\documentclass[aps,prd,showpacs,eqsecnum,floats,nofootinbib,superscriptaddress,preprint,tightenlines]{revtex4}

\usepackage[dvips]{graphicx}
\usepackage[mathscr]{eucal}
\usepackage[latin1]{inputenc}
\usepackage{amsmath}
\usepackage{amsfonts}
\usepackage{amssymb}
\usepackage{pictex}
\usepackage{epsfig}
\usepackage{colordvi}
\usepackage{color}

\def\be{\nopagebreak[3]\begin{equation}}
\def\ee{\end{equation}}
\def\ba{\nopagebreak[3]\begin{eqnarray}}
\def\ea{\end{eqnarray}}
\def\bas{\nopagebreak[3]\begin{eqnarray*}}
\def\eas{\end{eqnarray*}}

\def\d{{\rm d}}

\newcommand{\teta}{\rlap{\lower2ex\hbox{$\,\tilde{}$}}\eta{}}

\newcommand{\bi}{\begin{itemize}}
\newcommand{\ei}{\end{itemize}}

\newcommand{\mb}[1]{\mathbb{#1}}

\def\h{\hat }
\def\l{\lambda}
\def\lp{{\ell}_{\rm Pl}}
\def\bb{{\beta}}

\def\la{\langle}
\def\ra{\rangle}
\def\phib{\phi_{_{\rm b}}}
\def\phinfty{\phi \rightarrow \pm \infty}

\newcommand{\f}{\frac}

\usepackage{enumerate}

\usepackage{colordvi}
\newcounter{mnotecount}[section]

\newcommand{\comment}[1]{}

\def\f{\frac}

\def\t{\tilde}

\def\phy{\mathrm{phy}}

\begin{document}
\preprint{\vbox{\baselineskip=12pt \rightline{IGC-11/5-6}
}}

\title{Coherent Semiclassical States for 
Loop Quantum Cosmology}
\author{Alejandro Corichi}\email{corichi@matmor.unam.mx}
\affiliation{Instituto de Matem\'aticas,
Unidad Morelia, Universidad Nacional Aut\'onoma de
M\'exico, UNAM-Campus Morelia, A. Postal 61-3, Morelia, Michoac\'an 58090,
Mexico}
\affiliation{Center for Fundamental Theory, Institute for Gravitation and the Cosmos,
Pennsylvania State University, University Park
PA 16802, USA}
\author{Edison Montoya}
\email{eamonto@gmail.com}
 \affiliation{Instituto de F\'{\i}sica y
Matem\'aticas,  Universidad Michoacana de San Nicol\'as de
Hidalgo, Morelia, Michoac\'an, Mexico}
\affiliation{Instituto de Matem\'aticas,
Unidad Morelia, Universidad Nacional Aut\'onoma de M\'exico,
UNAM-Campus Morelia, A. Postal 61-3, Morelia, Michoac\'an 58090,
Mexico}

\begin{abstract}
The spatially flat Friedman-Robertson-Walker (FRW) cosmological model with a massless scalar field in loop quantum cosmology admits a description in terms of a completely solvable model.
This has been used to prove that: 
i) the quantum bounce that replaces the big bang singularity is generic; 
ii) there is an upper bound on the energy density for all states and 
iii) semiclassical states at late times had to be semiclassical before the bounce.
Here we consider a family of exact solutions to the theory, corresponding to generalized coherent Gaussian and squeezed states.
We analyze the behavior of basic physical observables and impose restrictions on the states based on physical considerations. 
These turn out to be enough to select, from all the generalized coherent states, those that behave semiclassical at late times.
We study then the properties of such states near the bounce where the most `quantum behavior' is expected. As it turns out, the states remain sharply peaked and semiclassical {\it at the bounce} and the dynamics is very well approximated by the `effective theory' throughout the time evolution. We compare the semiclassicality properties of squeezed states to those of the Gaussian semiclassical states and conclude that the Gaussians are better behaved. In particular, the asymmetry in the relative fluctuations before and after the bounce are negligible, thus ruling out claims of so called `cosmic forgetfulness'. 
\end{abstract}

\maketitle

\section{Introduction}

Loop quantum cosmology has provided in past years a 
useful framework to ask questions about the quantum nature
of the early universe \cite{lqc,AA}. Closely related to loop
quantum gravity \cite{lqg}, the formalism has shown that, for isotropic
models, the big bang singularity is replaced by a quantum
bounce \cite{aps0,aps,aps2}. These results have been obtained for
closed and open FRW cosmologies \cite{closed,open} with and without a cosmological
constant \cite{vp} for a massless scalar field, and recently extended to the
massive case \cite{massive}. Recently, some of these results have been extended to
anisotropic \cite{bianchiold,bianchiI,bianchiII,bianchiIX} and some inhomogeneous 
models \cite{gowdy}. In all these cases, quantum gravitational effects have been shown to 
exert a repulsive force and halt the collapse and launch an expanding 
superinflationary phase before the quantum gravity effects die out and
the universe then follows the standard general relativity (GR) dynamics. 

An exactly solvable model for the $k$=0 FRW with a massless scalar field
has allowed to prove analytically the generic nature of the quantum bounce \cite{slqc}.
In particular,  it was shown that the bounce occurs for all physical states, and the
energy density  possesses a supremum for physical states \cite{slqc}.
Regarding the fluctuations, it has been shown that they are bounded  across the bounce implying
that a semiclassical state at late times after the bounce had to come from a semiclassical state before the bounce
\cite{cs:prl}. 
Recently, Kaminski and Pawlowski have imposed stronger bounds on the asymmetry in
relative dispersions before and after the bounce \cite{kam:paw}, thus
very strongly ruling out some claims of loss of semiclassicality in
\cite{odc,harmonic}.
These robustness result have support on the fact that, within a large possibility of 
`loop quantizations', physical considerations select a unique consistent quantization 
\cite{cs:unique} (the one introduced in \cite{aps2}).

An interesting feature of this system is that one can  approximate very well the dynamics followed
by semiclassical states by an {\it effective description} \cite{aps2}. That is, the classical limit 
of the dynamics  is not that of GR + scalar field but rather is defined by an {\it effective Hamiltonian}.
There have been several approaches to reach this result. For instance, in \cite{vt} the author considered
{\it kinematical} coherent states and used them to compute the effective Hamiltonian as the expectation 
value of the quantum Hamiltonian constraint of LQC. More recently, the authors of \cite{ach} showed that, in the
path integral description of the model, the paths that contribute the most are those that satisfy the
effective equations and not those that follow the classical GR equations.

The purpose of this paper is to construct exact physical semiclassical states for the solvable model \cite{slqc}.
The strategy will be to define a class of coherent initial states and, 
using the analytical control we have on the space of
solutions to the quantum constraint equations, produce exact physical coherent states, much in the spirit of
\cite{semiclassical-coherent}. Then, we impose well motivated conditions on the behavior of the fluctuations 
of physical operators to select from all possible coherent states, those that exhibit a semiclassical behavior.
It turns out that these conditions are sufficient to fix enough parameters of the states and to have states that
are `peaked' on a given (physical) phase space point. Having found semiclassical states, we explore the behavior 
of these states near the bounce, where one expects the state to be most quantum. In particular we consider the relative 
fluctuations of volume and find that, for all states and for all times, the relative fluctuations are bounded by
their asymptotic value. In fact, the minimum value turns out to be precisely at the bounce. The expectation value
for volume and density on semiclassical states are very well approximated by the values given by the effective theory.
This leads us to conclude that the effective description is indeed a very good approximation to semiclassical {\it physical} states. A natural question is whether the class of states considered is the most general one. For that purpose, we generalize the class of Gaussian states to the so called {\it squeezed} states. The idea is to explore the parameter space near the Gaussian states --that are well behaved semiclassical states-- and see whether these squeezed states exhibit better semiclassical properties. 

Of particular interest in this regard is the asymmetry in volume fluctuations.
Claims of a loss of coherence, or `cosmic forgetfulness' arise due to results in an approximate framework \cite{odc} (motivated by, but not within LQC)  that suggest that the
fluctuations across the bounce might change significantly and spoil the semiclassical nature of the state \cite{odc,harmonic}. Even when
strong bounds against this possibility within LQC have already appeared \cite{kam:paw}, our {\it exact} formalism is particularly well suited for addressing this question. By imposing bounds on the total asymptotic dispersion in volume (at late or early times with respect to the bounce), we are able to bound how far we can be from Gaussian states and still have a semiclassical state. 
The range on parameter space allowed turns out to depend on the physical requirements for the state, in such a way that, for `large volume universes', it is severely constrained. This is turn sets very strong bounds on the
allowed asymmetry in volume fluctuations, consistent with \cite{kam:paw} and, in particular, disproving some of the statements of \cite{odc,harmonic} that claim the loss of coherence across the bounce for semiclassical states.

The structure of the paper is as follows. In Sec.~\ref{sec:2} we recall the solvable model in loop quantum cosmology together with the basic operators to be studied.
In Sec.~\ref{sec:3} we introduce the Gaussian states and compute expectation values and fluctuations for the
physical operators. Semiclassicality conditions are explored in Sec.~\ref{sec:4}, together with the behavior
of the states near the bounce. Squeezed states are introduced and compared to the Gaussian states in 
Sec.~\ref{sec:5}. We end with a discussion in Sec.\ref{sec:6}. There are four Appendices. In Appendix~\ref{app:a}
we study the symplectic reduction of the `effective theory' to isolate the physical degrees of freedom. In Appendix~\ref{app:b1} we recall some useful integrals and formulas. In Appendix~\ref{app:b} we present an analysis of the errors introduced by considering Gaussian states, showing that they are severely suppressed and justify our analysis in the main body. In Appendix~\ref{app:c} we present in a table a comparison of different quantities for two families of Gaussian states. Some of the results reported here were summarized in the short manuscript \cite{CM-letter}, so it is natural that there is some overlap of part of the material.

\section{Solvable Loop Quantum Cosmology (sLQC)}
\label{sec:2}

In this section we shall recall the model we are considering, namely,
the solvable $k$=0 FRW model coupled to a massless scalar field.
Our presentation will be self-contained and 
slightly different from that of \cite{slqc}.
In terms of the phase space variables used in loop quantum gravity
(LQG) \cite{lqg}, namely a connection $A^i_a$ and a densitized
triad $E^a_i$, the homogeneous gravitational sector can be expressed as,
\be A_a^i=c\,V_0^{-(1/3)}\,{}^{\rm o}\omega^i_a\quad ; \quad
E^a_i=p\, V_0^{-(2/3)}\,\sqrt{q_0}\,{}^{\rm o}e_i^a \ee
where $({}^{\rm o}\omega^i_a,{}^{\rm o}e_i^a)$ are a set of
orthonormal co-triads and triads compatible with the fiducial
(flat for $k=0$) metric ${}^{\rm o}q_{ab}$ and $V_o$ is the volume
of the fiducial cell, introduced to define the symplectic structure,
with respect to  ${}^{\rm o}q_{ab}$. The  phase space is
characterized by conjugate variables $(c,p)$ satisfying 
$\{c,p\}=(8\pi G\gamma)/3$.
%
where $\gamma$ $\approx 0.2375$ is the Barbero-Immirzi parameter
used in LQC. The triad $p$ is related to the physical volume $V$
of the fiducial cell,
as $|p| = V^{2/3} = V_o^{2/3} a^2$ where $a$ is the scale factor
and, on the space of classical solutions, $c = \gamma \dot a$. It is
convenient to introduce new variables \cite{slqc},
\be
  {\bb}:=\varepsilon\,c/p^{1/2}\qquad{\rm and}\qquad
\nu = \varepsilon\,p^{3/2}/(2 \pi \lp^2 \gamma) ~ \label{b_eq} \ee
where $\hbar\{\bb,\nu\} = 2$\footnote{Here we have adopted the
notation for the variable $\bb$ introduced in \cite{cs:unique} and used thereafter.}. 
Here $\varepsilon = \pm 1$ is the orientation of the triad with respect
to that of ${}^{\rm o}\omega^i_a$. The classical constraint, with
the choice $N=V$, then becomes
\be
 {\cal C}= \f{3}{8 \pi G \gamma^2}\,\bb^2 V^2 -  \f{p_\phi^2}{2} \approx 0
\ee
where the matter content consists of a massless scalar field $\phi$, with
canonical momenta $p_\phi$. Thus, the kinematical phase space  can be though
of as a four dimensional space with coordinates $(\bb,V,\phi,p_\phi)$. The
strategy for quantization is to define an operator associated to $\h {\cal C}$
and to look for states that are annihilated by the constraint operator.

In the $(\bb,\phi)$ representation, the operators basic are represented as
\ba
\hat{\bb}\cdot \chi(\bb,\phi)&=&\bb\,\chi(\bb,\phi)\quad ; \quad
\hat{V}\cdot  \chi(\bb,\phi) =
-i4\pi\gamma\lp^2\f{\partial}{\partial\bb}\, \chi(\bb,\phi) \, ; \\
\hat{p}_\phi \cdot \chi(\bb,\phi) &=& -i\hbar \f{\partial}{\partial\phi} \chi(\bb,\phi) \, .
\ea
However, the operator $\hat{\bb}$ is not well defined since
$\bb$ is periodic with period $2\pi/\l$, so it  has to be replaced by a well
defined operator $\hat{\tilde{\bb}}:=\sin(\lambda \bb)/\l$ (see \cite{slqc} for details).
 The quantum constraint of sLQC becomes then \cite{slqc},
\be \f{\partial^2}{\partial \phi^2}\cdot\chi(\bb,\phi)= \alpha^2
\,
\left(\frac{\sin(\lambda\bb)}{\lambda}\;\frac{\partial}{\partial
\bb} \right)^2 \cdot \chi(\bb,\phi)\label{const1} \ee
with $\bb\in (0,\pi/\lambda)$, $\alpha := \sqrt{12 \pi G}$. In the 
loop quantum cosmology literature, the value of $\lambda$ is chosen
such that $\lambda^2=\Delta$ corresponds to the minimum eigenvalue of the 
area operator in loop quantum gravity (corresponding to an edge
of `spin 1/2'). With this choice the free parameter $\lambda$ becomes
$\lambda^2 = 4 \sqrt{3} \pi \gamma \lp^2$. 
 By introducing
\be x = \alpha^{-1} \ln |\tan(\lambda \bb/2)| \label{x-eq} \ee the
quantum constraint can be rewritten in a Klein-Gordon form 
\be \label{kg-eq}
\partial_\phi^2 \, \chi(x,\phi) = \partial_x^2 \, \chi(x,\phi) ~.
\ee
As usual for this system, in its classical evolution
the scalar field $\phi$ is a monotonic function and 
can play the role of internal clock. In the quantum theory, 
one can also think of the evolution of the state with respect to
$\phi$.
A general solution $\chi(x,\phi)$ to Eq.~(\ref{kg-eq}) can be
decomposed in the left and right moving components:
\be\label{chi_eq} \chi = \chi_+ (\phi + x) + \chi_-(\phi - x) :=
\chi_+(x_+) + \chi_-(x_-) ~. \ee
The physical states that we shall consider are positive frequency 
solutions of ($\ref{kg-eq}$). Since there are no fermions in the model, the
orientations of the triad are indistinguishable and $\chi(x,\phi)$
satisfy the symmetry requirement $\chi(-x,\phi) = -\chi(x,\phi)$.
Thus, we can write
$\chi(x,\phi) =  (F(x_+) - F(x_-))/\sqrt{2}$, where $F$ is an
arbitrary `positive frequency solution'\footnote{To be precise, $F(x)$
is a positive momentum function, i.e. with a Fourier transform that has support
on the positive axis. With such a choice, the solution to the constraint equation
become of positive frequency.}.
 The physical inner product on solutions is given as \cite{slqc}
\ba
(\chi_1, \chi_2)_{\rm phy}
&=& -i\int_{\phi =\phi_0} [\bar\chi_1(x,\phi)\partial_\phi \chi_2(x,\phi)
-(\partial_\phi \bar\chi_1(x,\phi))\chi_2(x,\phi)] \, \d x\\
&=&i\int_{-\infty}^\infty [\partial_x \bar  F_1(x_+) F_2(x_+)
-\partial_x \bar  F_1(x_-) F_2(x_-)] \, \d x  ~. \label{inner-prod}
\ea
We can now compute the expectation values and fluctuations of fundamental
operator such as 
$\hat V|_{\phi{_o}}, \hat x$ and $ \hat p_\phi$, where $\hat x$ is related to
the operator $\hat \beta$.  For {\it any} state on the physical Hilbert
space the expectation value of the volume operator at `time $\phi$' is given by
\be
\la\hat{V}\ra_\phi :=
(\chi, \hat{V}|_\phi\chi)_{\rm phy}
 = 2\pi \gamma \lp^2 (\chi ,|\hat \nu| \chi)_{\rm phy}
\ee
where $|\hat \nu|$ is the absolute value operator obtained from
\be
\hat \nu=-\f{2\lambda}{\alpha}\cosh(\alpha x)i\partial_x \, .
\ee
Using the inner product \eqref{inner-prod} the expectation value of $|\hat \nu|$ is given by
\ba
(\chi,|\hat \nu| \chi)_{\rm phy}
&=&i\int_{-\infty}^\infty [\partial_x  \bar F(x_+)( \hat \nu F(x_+))
-\partial_x \bar F(x_-)(-\hat \nu F(x_-))] \, \d x \nonumber \\
&=& \f{2\lambda}{\alpha}\int_{-\infty}^\infty 
[\partial_x \bar F(x_+)\cosh(\alpha x)\partial_x F(x_+)
+\partial_x\bar F(x_-)\cosh(\alpha x)\partial_x F(x_-)]\, \d x \nonumber \\
&=& \f{4\lambda}{\alpha}\int_{-\infty}^\infty 
\left|\f{\d F}{\d x}\right|^2 \cosh(\alpha(x-\phi)) \, \d x \, .
\ea
Is easy to check that $|\hat \nu|=|\hat \nu|^\dagger$. From these
expressions one can find the expectation value of $\hat{V}^2$ and the
dispersion of the operator,
\begin{eqnarray}
\langle\hat{V}\rangle_\phi &=& V_+\,e^{\, \alpha \,\phi}
+V_-\,e^{-\alpha\,\phi} \label{v-exp} ~,\\
\langle\hat{V}^2\rangle_\phi~&=& \hskip-0.1cm
W_0+W_+\,e^{2\alpha\,\phi}
+W_-\,e^{-2\alpha\,\phi}\,, \label{fluct} \\
(\Delta\hat{V})_\phi^2&=&Y_0+Y_+\,e^{2\alpha\,\phi}
+Y_-\,e^{-2\alpha\,\phi} \label{fluct2} ~,
\end{eqnarray}
with $V_{\pm}$,  $W_0$, $W_{\pm}$, $Y_0$ and $Y_\pm$ being  real
and positive, given by 
\bas
V_{\pm} &= &\frac{4 \pi \gamma \lp^2
\lambda}{\alpha}\int \left|\frac{\d F}{\d
x}\right|^2\,e^{\mp\alpha\, x} \d x \label{const5} ~,\nonumber \\
W_0 &=& \f{i 2\pi \gamma^2\lp^4\lambda^2}{3G}  \int
\left[\f{\d^2\bar{F}}{\d x^2}\,\f{\d F}{\d x} - \f{\d \bar{F}}{\d
x}\,\f{\d^2 F}{\d x^2}\right] \d x ~,\nonumber \\
\label{w-eq} W_\pm &=&\f{i\pi \gamma^2\lp^4\lambda^2}{3G} \int
e^{\mp 2\alpha\,x} \left[\f{\d^2\bar{F}}{\d x^2}\,\f{\d F}{\d
x} - \f{\d \bar{F}}{\d x}\,\f{\d^2 F}{\d x^2}\right] \d x~,
\eas
for normalized states. Here $Y_0 = W_0 - 2 V_+ V_-$ and $Y_\pm
= W_\pm - V_\pm^2$. Note that the expressions  for $W_0$ and $W_\pm$ correct 
those found in  \cite{cs:prl}.

From (\ref{v-exp}), it follows that the expectation value of the volume $\hat
V|_\phi$ is large at both very early and late times and  has a
non-zero global minimum $V_{\mathrm{min}} = 2 (V_+
V_-)^{1/2}/||\chi||^2$. The {\it bounce} ocurrs at time
$\phi_b^V = (2 \alpha)^{-1} \ln(V_-/V_+)$ \cite{slqc}. Around
$\phi = \phi_b^V$, the expectation value of the volume $\langle
\hat V\rangle_\phi$ is symmetric. Similarly, $\langle \hat V^2
\rangle_\phi$ is symmetric across the value $\phi_b^{V^2}=(4
\alpha)^{-1}\ln(W_-/W_+)$ for the scalar field. A trivial
observation is that if $\phi_b^V = \phi_b^{V^2}$, the difference
in the asymptotic values of the relative fluctuation
\be\label{D-eq}
D:=\lim_{\phi\to\infty}\left[\left(\frac{(\Delta\hat{V})}{\langle\hat{V}\rangle}\right)_{\phi}^2
-
\left(\frac{(\Delta\hat{V})}{\langle\hat{V}\rangle}\right)_{-\phi}^2\right]
= \frac{W_+}{V^2_+}-\frac{W_-}{V^2_-} \ee
vanishes. It should be noted that this quantity quantifies the change in
semiclassicality across the bounce as pointed out in \cite{cs:prl}.

The other observable that one might want to consider as fundamental,
is an observable naturally conjugate to $p_\phi$. Given that we are
considering physical states, this does not correspond to the
operator $\phi$ as happens in the kinematical setting. Instead, as
illustrated in the Appendix~\ref{app:a}, the quantity that is conjugate to
$p_\phi$, in the reduced physical model, corresponds to the quantity
$x(\bb)$. Thus, it is natural to consider the family of operators
$\hat{x}_\phi$, corresponding to the observable `$x$ at  time
$\phi$'.
The expectation value of $\hat x$ is defined as,
\be
(\chi_1,\hat x \chi_2)_{\rm phy}
=\f{1}{2}[(|\hat x|\chi_1,\chi_2)_{\rm phy}+(\chi_1,|\hat x|\chi_2)_{\rm phy}] \label{x_sym}
\ee
where $|\hat x|F(x_+)=xF(x_+)$ and $|\hat x|F(x_-)=-xF(x_-)$. We have included the two terms in the
definition \eqref{x_sym} because $|\hat x|$ is not symmetric.
We can now compute the expectation values and fluctuations of
$\hat x$ for {\it any} state of the physical Hilbert space
\ba
\la \hat x\ra_\phi &=& \t{X}_1-\phi \, , \label{x-ope}\\
\la \hat x^2\ra_\phi &=& \t X_2 -2\phi \t{X}_1+ \phi^2 \, ,\\
(\Delta \hat x) &=& \t X_2 - \t{X}_1^2  \, ,
\ea
with $\t X_1$ and $\t X_2$ real (state dependent) constants, given by
\ba
\t X_1 &=&-2i\int_{-\infty}^\infty \bar F(x)x\partial_{x} F(x) \d x
-i\int_{-\infty}^\infty |F(x)|^2 \d x \label{X_1}\\
\t X_2 &=&-2i\int_{-\infty}^\infty \bar F(x) x^2 \partial_x F(x) \d x 
- 2i\int_{-\infty}^\infty  |F(x)|^2 x \d x', , \label{X_2}
\ea
for normalized states. Is important to note that $\la \hat x\ra_\phi \propto \phi$,
$\la \hat x^2\ra_\phi \propto \phi^2$ but the fluctuation $\Delta\hat x$ is independent of
$\phi$! This property indeed gives support to the proposal that $\hat x$ be the conjugate of $\hat p_\phi$. 
Finally, we compute the expectation values for $p_\phi$, a Dirac observable as
\ba
\hat p_\phi=-i\hbar \partial_\phi & \longrightarrow & 
\la \hat p_\phi \ra = 2 \hbar \int_{-\infty}^\infty\left|\f{\d F}{\d x}\right|^2\d x \\
\hat p_\phi^2 =-\hbar^2 \partial_\phi^2 & \longrightarrow &
\langle \hat p_\phi^2 \rangle=- 2 \hbar^2 i \int_{-\infty}^\infty
\f{\d \bar F}{\d x}\f{\d^2 F}{\d x^2}\d x', ,
\ea
for normalized states.
In the following sections, it will be convenient to consider a general physical state written as, 
\be \chi(x,\phi) = \int_0^\infty
\tilde{F}(k)\,e^{-ik(\phi+x)} \,\d k-
\int_0^\infty \tilde{F}(k)\,e^{-ik(\phi-x)} \,\d k 
\ee 
where the Fourier transform
$\tilde{F}(k)$ contains all the information of the state. Positive frequency
solutions to Eq.~(\ref{kg-eq}) means that $\tilde{F}(k)$ has support on positive
$k$'s only. Furthermore, we
can consider initial states, for an `initial time' $\phi=0$, in order to
compute some of the quantities that determine the expectation value of 
relevant operators. It then suffices to specify
$\tilde{F}(k)$ and from there to compute all the 
quantities $V_\pm, W_0, W_\pm, \t X_1, \t X_2, \la \hat p_\phi \ra$
and $\la \hat p_\phi^2 \ra$.


\section{Gaussian Initial States}
\label{sec:3}

In this section we shall consider physical states with initial
states given {\it generalized Gaussian states}, defined 
by functions $\t F(k)$ of the form:
\be \label{coh_state}
\t F(k)=\left\{
\begin{array}{rl}
k^n e^{-(k-k_0)^2 /\sigma^2}e^{ipk},& \mbox{ for } k>0 ,\\[1ex]
0,& \mbox{ for } k\le 0 ,
\end{array}
\right.
\ee
with $\sigma>0, k_0>0$, $ n=0,1,2,...$.
That is, we are choosing Gaussian states centered around the point
$k_0$, with `dispersion' given by $\sigma$.

The physical states one has to consider are
vanishing in the negative $k$ axis, to
have positive frequency solutions. In order to gain analytical control
over all the quantities at hand, we
shall in what follows consider instead
the Gaussian states for all values of $k$.
Thus, the quantities we shall compute are an approximation to the
real quantities but, as we shall argue in detail, this approximation 
is justified when the values of
$k_0$ and $\sigma$ are such that the wave function has a negligible contribution
from the negative axis. A detailed analysis of the errors
introduced by this simplification can be found in the Appendix~\ref{app:b}.
Therefore, from now on, in order to have states that approximate
positive frequency solutions, we shall impose the condition $k_0\gg \sigma$.
Further consistency conditions on $\sigma$, motivated by physical
considerations will be derived below.

The norm of the state in $k$ space is
\[
(\t\chi,\t\chi)_{\rm phy}=-2i \int_{0}^\infty \bar{\t F} (k) ik \t F(k)
\d k =2\int_{0}^\infty k |\t F(k)|^2 \d k .
\]
If we take the initial states in the $k$-space as in
(\ref{coh_state}), the norm  is,
\[
(\t\chi,\t\chi)_{\rm phy}=2\int_{0}^\infty k k^{2n}e^{-2(k-k_0)^2
/\sigma^2} \d k \, ,
\]
which can be approximated by $(\t\chi,\t\chi)_{\rm
phy}=2\int_{-\infty}^\infty k k^{2n}e^{-2(k-k_0)^2 /\sigma^2} \d
k$ (see Appendix~\ref{app:b} for a justification of this approximation).

With the change of variables $u=\f{\sqrt{2}}{\sigma}(k-k_0)$ this last
integral takes the form
\be \label{norm_int}
\|\t\chi\|_{\rm
phy}^2=2\left[\f{\sigma}{\sqrt{2}}\right]^{2n+2}
\int_{-\infty}^\infty \left[u+\f{\sqrt{2}
k_0}{\sigma}\right]^{2n+1} e^{-u^2} \d u\, .
\ee
For concreteness, we shall consider the state corresponding to $n=0$. In this case
\ba \|\t\chi\|_{\rm
phy}^2&=&2\left[\f{\sigma}{\sqrt{2}}\right]^{2}
\int_{-\infty}^\infty \left[u+\f{\sqrt{2} k_0}{\sigma}\right] e^{-u^2} \d u 
=\sqrt{2\pi}\, k_0\sigma \, . \label{norm0} 
\ea
%

The normalized $n$=0 states can be used to compute explicitly the
expectations values and fluctuations of several physically
interesting operators. The corresponding results for 
 $n$=1 are summarized in the Appendix~\ref{app:c}.  

\subsection{Expectation Values for Basic Observables}

Let us explore the basic observables for the Gaussian states in order
to gain a better understanding of the free parameters $(k_o,\sigma,p,n)$ that
characterize the states. The first observable we shall consider is
$p_\phi$, that is a Dirac observable (and therefore a `constant of
the motion'). The operator is represented as
$\hat{p}_\phi\cdot\chi=-i\hbar\f{\partial}{\partial\phi}\cdot\chi$,
and its expectation value is thus given by
\[
\langle \hat p_\phi \rangle=\f{2 \hbar}{\|\t\chi\|^2_\phy}
\int_{-\infty}^\infty\left|\f{\d F}{\d x}\right|^2\d x', .
\]
Using Parseval's theorem we get
\be \langle \hat p_\phi \rangle 
= \f{2 \hbar}{\|\t\chi\|^2_\phy} \int_{-\infty}^\infty k^2 |\t F(k)|^2 \d k . 
\ee
If we take our initial states 
$\t F(k)=k^n e^{-(k-k_0)^2 /\sigma^2}e^{ipk}$ 
then the integral takes the form \be \label{Pphi} \langle \hat p_\phi
\rangle = \f{2 \hbar}{\|\t\chi\|^2_\phy} \int_{-\infty}^\infty
k^{2n+2} e^{-2(k-k_0)^2 /\sigma^2} \d k \, .\ee
With the change of variables $u=\f{\sqrt{2}}{\sigma}(k-k_0)$, 
the expectation value (\ref{Pphi}) takes the form
\be
 \langle \hat p_\phi \rangle   = \f{2 \hbar}{\|\t\chi\|^2_\phy} \left[
\f{\sigma}{\sqrt{2}}\right]^{2n+3} \int_{-\infty}^\infty \left[
u+\f{\sqrt{2}k_0}{\sigma} \right]^{2n+2} e^{-u^2} \d u  \, .
\ee
In the case of pure Gaussian states, namely for $n=0$, we have
\be \langle \hat p_\phi \rangle = 
= \hbar k_0 \left[1 +\f{\sigma^2}{4 k_0^2} \right]  \, .\ee
%
%
This is telling us that, for the approximation we are taking, namely 
$k_0 \gg \sigma$, then $\langle \hat p_\phi \rangle \approx \hbar k_0$.
Therefore, as expected, the parameter $k_0$ is giving us a good
measure of the expectation value of $p_\phi$ which can be regarded as
the conjugate variable to $x$, on the reduced phase space of the system
(See Appendix~\ref{app:a}). Let us now compute the
fluctuations of this operator. Let us start by computing
\be
\langle \hat p_\phi^2 \rangle  = - \f{2 \hbar^2
i}{\|\chi\|^2_\phy} \int_{-\infty}^\infty \f{\d \bar{F}}{\d x}\f{\d^2
F}{\d x^2}\d x =
 \f{2\hbar^2}{\|\chi\|^2_\phy}
\int_{-\infty}^\infty k^3 |\t F(k)|^2 \d k', .
\label{parsebal2}
\ee
For our generalized Gaussian states the integral takes the form
\be \label{Pphi2} \langle \hat p_\phi^2 \rangle = \f{2 \hbar^2
i}{\|\chi\|^2_\phy} \int_{-\infty}^\infty k^{2n+3} e^{-2(k-k_0)^2
/\sigma^2} \d k \, .\ee
Which can be rewritten as,
\[
\langle \hat  p_\phi^2 \rangle=  \f{2 \hbar^2 i}{\|\chi\|^2_\phy}
\left[ \f{\sigma}{\sqrt{2}}\right]^{2n+4} \int_{-\infty}^\infty
\left[ u+\f{\sqrt{2}k_0}{\sigma} \right]^{2n+3} e^{-u^2}\, \d u\, .
\]
Taking $n=0$, 
\be \langle \hat p_\phi^2 \rangle  = 
= \hbar^2 k_0^2\left[1 +\f{3\sigma^2}{4 k_0^2} \right]\, . \label{p2-result}
 \ee
Note that for large $k_0$, that is for those states satisfying
$k_0 \gg \sigma$, then $\langle \hat p_\phi^2 \ra \approx \hbar^2 k_0^2 $.

\noindent Let us now compute the dispersion of the observable
$\hat p_\phi$, $(\Delta \hat p_\phi)^2=\langle \hat p_\phi^2 \rangle-\langle
\hat p_\phi \rangle^2$.
For the $n=0$ case we have
\ba
(\Delta \hat p_\phi)^2&=&\hbar^2 k_0^2\left[1 +\f{3\sigma^2}{4
k_0^2} \right]
-\hbar^2 k_0^2\left[1 +\f{\sigma^2}{4 k_0^2} \right]^2 \nonumber \\
&=& \f{\hbar^2\sigma^2}{4}\left[1 -\f{\sigma^2}{4 k_0^2}\right]
\ea
which is a constant. Note that for $k_0 \gg \sigma$, the
dispersion $\Delta \hat{p}_\phi\approx \f{\hbar \sigma}{2}$.
This is telling us that the parameter $\sigma$ has the
interpretation one might have expected as the dispersion of the
observable associated with the variable $k$, which in this case
corresponds to $p_\phi$.

Recall that the quantity that is conjugate to
$p_\phi$, in the reduced model, corresponds to
$x(\bb)$. Thus, it is natural to consider the family of operators
$\hat{x}_\phi$, corresponding to the observable `$x$ at the time
$\phi$'. It is most natural to define the symmetric operator
acting on initial states $\t F$, in the $k$-representation the operator
defined in Eq.~\eqref{x-ope} can be written as:
\be
 \hat{x} \t F:=\left(\f{i\partial}{\partial k} +
 \f{i}{2k}-\phi \right)\cdot \t F \, .\label{x-operator}
\ee
The first two terms arise when Eq. \eqref{X_1} is replaced into Eq. \eqref{x-ope}.
It is now straightforward to compute the expectation value of the
$\hat{x}$ operator on the generalized Gaussian states. It is easy
to see that, for all values of $n$,
\be
 \label{x_0} \la \hat x \ra= -p -\phi\, ,
\ee
which confirms the expectation that the parameter $-p$ represents
the point in $x$ space, where the Gaussian is peaked (when $\phi=0$). 
This also corresponds to the classical dynamics (as given by the effective
theory) for the variable $x$ (with our choice $p_\phi > 0$). Furthermore,
one can try to find the expectation value of the operator
\[\hat{x}^2=\left(\f{i\partial}{\partial k} +\f{i}{2k} -\phi\right)^2 =
-\f{\partial^2}{\partial k^2} -\f{1}{k}\f{\partial}{\partial k}+\f{1}{4k^2}
-2 \phi \left(\f{i\partial}{\partial k} + \f{i}{2k}\right) + \phi^2\, .
\]
Here it is important to note two important issues that arise when
considering the operator defined by (\ref{x-operator}). The first one is that it
involves a derivative operator and the second is the factor $1/k$
in the operator. The derivative term implies that one needs to
consider carefully the boundary conditions at $k=0$, so that the
boundary terms (when integrating by parts) does not contribute. In
particular, we should have, for the $\hat{x}$ operator, for
states defined by functions $f(k)$ and $g(k)$, that
$\left.(\bar{g}\,f\,k)\right|_{k=0}=0$. This condition is, of
course, satisfied by all of our states $n\geq 0$. The term $1/k$
imposes some fall-off conditions at the origin ($k=0$) for the
integrals to be finite. For instance, even when the expectation
value of $\hat{x}$ is well defined in the $n=0$ Gaussians, these
states are no longer in the domain of the operator. The state that
results, when acting with the operator $\hat{x}$ on a $n=0$ state
is not normalizable, and therefore, does not belong to the
physical Hilbert space. We have to conclude that, in order to have
the operator  $\hat{x}$ in the set of observables we want to
consider, we have to exclude the $n=0$ states and consider only
$n\geq 1$.
In the $n=1$ case, it is straightforward (if lengthy) to compute
the expectation value of the operator,
 \be \label{x2_0}
  \la \hat x^2
\ra_\phi = p^2+\f{1}{\sigma^2}\,\f{\left(1+\f{3\sigma^2}{2k_0^2} \right)}
{\left(1 +\f{3\sigma^2}{4k_0^2} \right)}
+ 2 p\phi  +\phi^2
 \ee
from which the fluctuations of the operator becomes,
\be \label{deltax} (\Delta \hat x)^2
= \la \hat x^2 \ra_\phi - \la \hat x\ra^2_\phi
=\f{1}{\sigma^2}\,\f{\left(1+\f{3\sigma^2}{2k_0^2} \right)}
{\left(1 +\f{3\sigma^2}{4k_0^2} \right)}
 \ee
which also tells us that the parameter $\sigma$ has the expected
interpretation of providing the inverse of the dispersion for the
observable $x$. This is realized when $k_0\gg\sigma$ and therefore the second term 
in the previous expression is very close to 1.

We can now examine the uncertainty relations for the observables
$\hat{p}_\phi$ and $\hat{x}$ (for $n=1$):
\be (\Delta \hat p_\phi)^2(\Delta \hat x)^2
= \f{\hbar^2}{4}\; \f{\left(1+\f{9\sigma^4}{16k_0^4}+
\f{3\sigma^2}{4k_0^2}-\f{9\sigma^6}{64k_0^6}\right)
\left(1+\f{3\sigma^2}{2k_0^2} \right)}
{\left(1+\f{3\sigma^2}{4k_0^2}\right)^3}
 \ee
We note that, if $k_0\gg\sigma$, then the uncertainty relations
$$
(\Delta \hat p_\phi)(\Delta \hat x) \geq \f{\hbar}{2}
$$
are very close to being saturated. As we shall see in the next
section, these conditions are necessary for the generalized
Gaussian states to represent acceptable semiclassical states.

\subsection{Volume Operator}

Let us start by considering the operator $\hat V|_{\phi_0}$
corresponding to the volume  at `time' $\phi_0$. The expectation
value of this operator is given by Eq. (\ref{v-exp}) so it
suffices to compute the constants $V_{\pm}$ given by 
\bas
V_{\pm} &= &\frac{\kappa}{\alpha \|\chi\|^2_\phy}
\int \left|\frac{\d F}{\d x}\right|^2\,e^{\mp\alpha\, x} \d x \, ,
\hspace{0.5cm} \mbox{with} \hspace{0.5cm} \kappa = 4 \pi \gamma \lp^2 \lambda 
\eas
which can be written as,
\ba 
V_{\pm}&=&\f{\t V_\pm}{\|\chi\|^2_\phy}=\f{1}{ \|\chi\|^2_\phy}\frac{\kappa}{\alpha}
\int k(k\mp i\alpha )\bar {\t F} (k) \t F(k\mp i\alpha )\,\d k \, .
\ea
If we take the state as $\t F(k)=k^n e^{-(k-k_0)^2 /\sigma^2}e^{ipk}$,
the integral is
\[
\t V_\pm=\f{\kappa}{\alpha} \int_{-\infty}^\infty k k^n
e^{-(k-k_0)^2 /\sigma^2}e^{-ipk} [k\mp i\alpha][k\mp i\alpha]^n
e^{-(k\mp i\alpha-k_0)^2 /\sigma^2}e^{ip(k\mp i\alpha)} \d k \, .
\]
This can be rewritten as,
\be \label{vol1} \t V_\pm=\f{\kappa}{\alpha} e^{\pm
p\alpha}\int_{-\infty}^\infty [k(k\mp i\alpha)]^{n+1}
e^{-(k-k_0)^2 /\sigma^2}e^{-(k\mp i\alpha-k_0)^2 /\sigma^2} \d k
\, .\ee
It is straightforward to see that, with the change of variable
$u=\sqrt{2}k-\sqrt{2}k_0\mp\f{i\alpha}{\sqrt{2}}$,
the integral  (\ref{vol1}) can be written as
\[
\t V_\pm=\f{\kappa}{\sqrt{2}\alpha}\left(\f{1}{4}\right)^{n+1}
e^{\pm p\alpha} e^{\alpha^2/2\sigma^2}\int_{-\infty}^\infty
\left[2u^2+4\sqrt{2}k_0 u+4k_0^2 +\alpha^2 \right]^{n+1}
e^{-u^2/\sigma^2}\d u\, .
\]
For the $n=0$ case we have,
\be\label{vol2} \t
V_\pm=\f{\kappa}{\alpha\sqrt{2}}\left(\f{1}{4}\right) e^{\pm
p\alpha} e^{\alpha^2/2\sigma^2}\int_{-\infty}^\infty
\left[2u^2+4\sqrt{2}k_0 u+4k_0^2 +\alpha^2 \right]
e^{-u^2/\sigma^2}\d u\, . \ee
%
%
From which we get
\[
\t V_\pm=\f{\kappa}{\alpha\,4\sqrt{2}}\; e^{\pm p\alpha}\,
e^{\alpha^2/2\sigma^2}
\sqrt{\pi}(\sigma^3+4k_0^2\sigma+\alpha^2\sigma )\, .
\]\, 
If we take now a normalized states using Eq.~(\ref{norm0}), the result
is
\be \label{Vpm_0} V_\pm=\f{\t V_\pm}{\|\t\chi\|_{\rm phy}^2}
=\f{\pi \gamma \lp^2 \lambda}{2\alpha }\; e^{\pm p\alpha }\,
e^{\alpha ^2/2\sigma^2} \left[4k_0 +\f{\sigma^2 +\alpha ^2}{k_0}
\right]\, . 
\ee
In order to calculate the dispersion of the volume operator, it
is necessary to calculate the three integrals $W_0,W_+,W_-$. Let
us start by computing the quantity $W_0$.
%
\be \label{W0_int}
W_0=\f{i\alpha_2}{\|\chi\|^2_\phy}\int_{-\infty}^\infty\left[\f{\d^2 \bar F}{\d x^2}
\f{\d F}{\d x}-\f{\d \bar F}{\d x}\f{\d^2 F}{\d x^2} \right]\d x
,\ \mbox{with }\ \alpha_2=\f{2\pi \gamma^2 \lp^4 \lambda^2}{3 G}\, ,
\ee
which can be written as
\be
W_0=\f{\t W_0}{\|\chi\|^2_\phy}
 =  \f{2\alpha_2}{\|\chi\|^2_\phy} \int_{-\infty}^\infty k^3|\t F(k)|^2 \d k\, .
\ee
This is the same integral that the one in Eq. \eqref{parsebal2}. Then, 
for a normalized Gaussian state with  $n=0$ we only
need to replace $\hbar^2$ by $\alpha_2$ into Eq. \eqref{p2-result} to obtain that
\be \label{W0_0}
W_0=\alpha_2 k_0^2\left[1 +\f{3\sigma^2}{4 k_0^2}\right]\, .
\ee
Let us now compute the quantities $W_\pm=\tilde{W}_\pm/\|\chi\|^2_\phy$, with
\be \label{Wpm_int}
\tilde{W}_\pm =i\alpha_3\int_{-\infty}^\infty e^{\mp 2\alpha x}\left[\f{\d^2 \bar F}{\d x^2}
\f{\d F}{\d x}-\f{\d \bar F}{\d x}\f{\d^2 F}{\d x^2} \right]\d x
,\ \mbox{with }\ \alpha_3=\f{\pi \gamma^2 \lp^4 \lambda^2}{3 G} \, ,
\ee
that can be written as,
\be
\tilde{W}_\pm =
\alpha_3\int_{-\infty}^\infty k^2
\left[\bar{\t F}(k)(k\mp 2i\alpha) \t F(k\mp 2i\alpha)
 +(k\pm 2i\alpha)\bar{\t F}(k\mp 2i\alpha)\t F(k)\right] \d k \, .\\
\ee
If we take the initial states
$\t F(k)=k^n e^{-(k-k_0)^2 /\sigma^2}e^{ipk}$,
the integral is
\ba
\t W_\pm 
&=&\alpha_3 e^{\pm 2 p\alpha}\int_{-\infty}^\infty k^{n+2} e^{-(k-k_0)^2 /\sigma^2} \nonumber \\
&&[ (k\mp 2i\alpha)^{n+1} e^{-(k\mp 2i\alpha-k_0)^2/\sigma^2}
+(k\pm 2i\alpha)^{n+1}e^{-(k\pm 2i\alpha-k_0)^{2}/\sigma^2}] \d k
\ea
In order to do the integral
\[
I(\pm)=\int_{-\infty}^\infty  k^{n+2} e^{-(k-k_0)^2 /\sigma^2}
(k\pm 2i\alpha)^{n+1}e^{-(k\pm 2i\alpha-k_0)^{2}/\sigma^2} \d k\, .
\]
It is straightforward to show that
with the change of variables
$u=\sqrt{2}(k-k_0\pm i\alpha )$, we get
%
\ba
I(\pm)&=&\int_{-\infty}^\infty  \left(\f{u}{\sqrt{2}}+k_0\mp i\alpha\right)^{n+2}
\left(\f{u}{\sqrt{2}}+k_0\mp i\alpha \pm 2i\alpha\right)^{n+1}
e^{-u^2 /\sigma^2} e^{2\alpha^{2}/\sigma^2} \f{\d u}{\sqrt{2}} \nonumber \\
&=&  \f{e^{2\alpha^{2}/\sigma^2}}{\sqrt{2}}\int_{-\infty}^\infty
 \left[\left(\f{u}{\sqrt{2}}+k_0\right)^2 + \alpha^2\right]^{n+1}
\left(\f{u}{\sqrt{2}}+k_0\mp i\alpha\right)
e^{-u^2 /\sigma^2} {\d u}\, .
\ea
The real part of $I(\pm)$ is
\be \label{real_int}
\mb{R}{\rm e}[I(\pm)]= \f{e^{2\alpha^{2}/\sigma^2}}{\sqrt{2}}\int_{-\infty}^\infty
 \left[\left(\f{u}{\sqrt{2}}+k_0\right)^2 + \alpha^2\right]^{n+1}
\left(\f{u}{\sqrt{2}}+k_0\right)
e^{-u^2 /\sigma^2} {\d u}\, .
\ee
Then
$\mb{R}{\rm e}[I(+)]=\mb{R}{\rm e}[I(-)]$, and the imaginary part of $I(\pm)$ is
\[
\mb{I}{\rm m}[I(\pm)]=\mp\alpha \f{e^{2\alpha^{2}/\sigma^2}}{\sqrt{2}}\int_{-\infty}^\infty
 \left[\left(\f{u}{\sqrt{2}}+k_0\right)^2 + \alpha^2\right]^{n+1}
e^{-u^2 /\sigma^2} {\d u}\, ,\\
\]
therefore, $\mb{I}{\rm m}[I(+)]=-\mb{I}{\rm m}[I(-)]$.
Using the real and imaginary parts, we have that the integral $\t W_\pm$ takes the form
\be 
\t W_\pm=\alpha_3 e^{\pm 2p\alpha}[I(+)+I(-)]
=2\alpha_3 e^{\pm 2p\alpha}\mb{R}{\rm e}[I(+)]\, .
\ee
This tells us that $\t W_\pm$ is real valued (as expected) and
that it is only necessary to calculate  the integral
(\ref{real_int}).
Taking $n=0$ the integral (\ref{real_int}) takes the form
\ba
\mb{R}{\rm e}[I(\pm)]&=& \f{e^{2\alpha^{2}/\sigma^2}}{\sqrt{2}}\int_{-\infty}^\infty
 \left[\left(\f{u}{\sqrt{2}}+k_0\right)^2 + \alpha^2\right]
\left(\f{u}{\sqrt{2}}+k_0\right)
e^{-u^2 /\sigma^2} {\d u}  \nonumber \\
&=& \f{e^{2\alpha^{2}/\sigma^2}}{\sqrt{2}}\int
 \left(\f{u^3}{2^{2/3}}+3\f{u^2}{2}k_0+ \f{u}{\sqrt{2}}(3k_0^2+\alpha^2)
+(k_0^3 +\alpha^2k_0)\right)e^{-u^2 /\sigma^2} {\d u}\, .  \nonumber 
\ea
Using the integrals from Appendix~\ref{app:b1}, the integral is
\ba
\mb{R}{\rm e}[I(\pm)]
&=& \f{e^{2\alpha^{2}/\sigma^2}\sqrt{\pi}}{\sqrt{2}}
 \left(\f{3}{4}k_0\sigma^3
+k_0^3\sigma +\alpha^2k_0\sigma \right)\, ,
\ea
from which we have that,
\be
\t W_\pm  = \alpha_3 e^{\pm 2p\alpha}{e^{2\alpha^{2}/\sigma^2}\sqrt{2\pi}}
 \left(\f{3}{4}k_0\sigma^3+k_0^3\sigma +\alpha^2k_0\sigma \right)\, .
\ee
Including now the normalization of the state we have
\be
\label{Wpm_0} W_\pm  =  \f{\pi \gamma^2 \lp^4 \lambda^2}{3 G}\;
e^{\pm 2p\alpha }\,e^{2\alpha ^{2}/\sigma^2}\,
 \left[k_0^2 +\f{3}{4}\sigma^2+\alpha ^2 \right]\, .
\ee
From the last equation and Eq.~(\ref{Vpm_0})  we can observe that
the difference in the asymptotic relative fluctuations after and before the bounce, 
as defined by Eq.~\eqref{D-eq}, vanishes. That is,  $D=0$.
It is important to note that this is a general result for any
Gaussian state of the form of Eq. (\ref{coh_state}), for all
values of $n=0,1,2,...$, as was  expected to happen \cite{cs:prl}.

\section{Semiclassicality Conditions}
\label{sec:4}

So far we have considered generalized Gaussian states as initial
states for physical states in the exactly solvable $k$=0 LQC model
with a massless scalar field. We have seen that the parameters
that define the states, in the pure Gaussian case with $n=0$, have
the expected interpretation: The parameter $k_0$ is related to the
expectation value of $\hat{p}_\phi$, $-p$ to the expectation value of
$\hat{x}$, and $\hbar\sigma/2$  to the dispersion of $p_\phi$. 
The only condition that we have imposed so far is $k_0 \gg \sigma$,
which guaranties the validity of our approximation.
We have also seen that, for states which satisfy this consistency
condition, there are
at least two results that follow. First,  it warranties
that the relative dispersion of $p_\phi$ is small. Second,
this conditions implies that the uncertainty relations become
saturated in the sense that $(\Delta \hat{p}_\phi)(\Delta\hat{x})
\approx\hbar/2$. 

From this perspective, it could seem that any value of $k_0$ and $\sigma$,
provided they satisfy $k_0 \gg\sigma$, might be acceptable to
define semiclassical states. 
The purpose of this section is to explore this issue further and
answer the following question: Are there more stringent conditions
that one must impose in order to have semiclassical states that
will further restrict the possible values of $k_0$ and $\sigma$?
As we shall see, the answer is in the affirmative.

\subsection{Asymptotic Volume}

\begin{figure}[tbh!]
\includegraphics[scale=0.65]{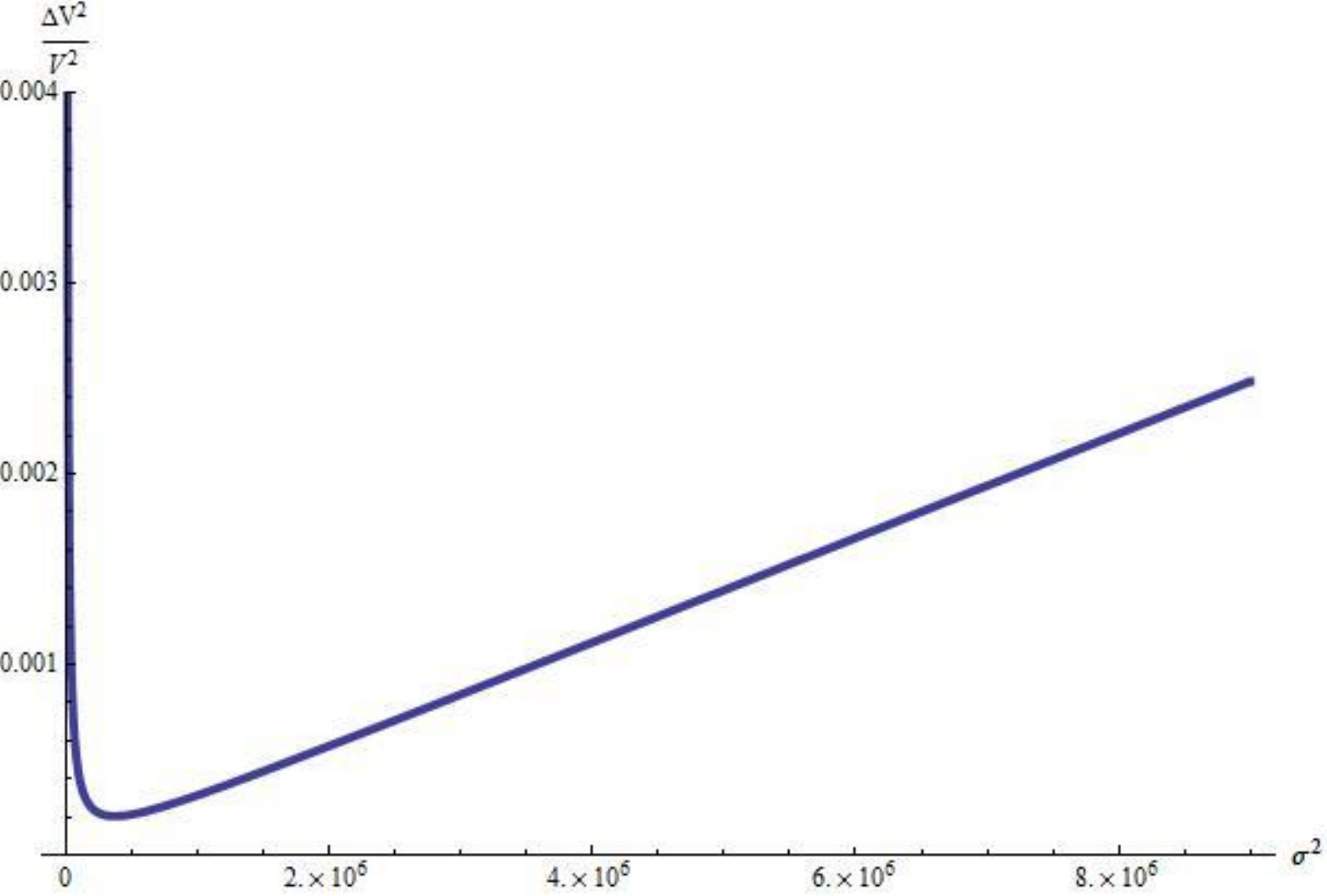}
\caption{The asymptotic relative fluctuation $\Delta_\pm$ is plotted as a function
        of $\sigma^2$ with $k_0=30000$ in Planck units, where
        $\alpha\approx 6.14$. Here we have chosen $n=0$ and $p=0$.
        Two features characterize the behavior of the function near
        the minimum $\t\sigma$. The first one is that for $\sigma^2<\t\sigma^2$ the function is
        exponential; the second one is that for $\sigma^2>\t\sigma^2$ the function has a
        polynomial behavior.
}\label{fig:1.1}
\end{figure}

In order to answer this question we shall first consider the volume
operator. Since all physical states have the property that the
expectation value of the volume operator $\hat{V}|_\phi$ at time
$\phi$ follow the {\it same} functional form and, therefore,
follow for large $\phi$ the same dynamics of the classical
dynamics (i.e. Einstein's equations) $V(\phi) \sim
e^{\pm\alpha\phi}$, we need more criteria to select those states
that are semiclassical. The obvious strategy is to consider the
state's relative dispersion
$\left((\Delta\hat{V})/\langle\hat{V}\rangle\right)_{\phi}^2$ of
the volume operator at time $\phi$. One expects that semiclassical
states will have a very small relative dispersion `at late times'
when the dynamics approaches the classical dynamics. It is then
natural to consider the asymptotic relative dispersions given by
\be
 \Delta_\pm:=\lim_{\phi\to\pm\infty}
 \left(\f{\Delta\hat{V}}{\langle\hat{V}\rangle}\right)_{\phi}^2=
 \f{W_\pm}{V_\pm^2}-1\, .
\ee
It is straightforward to find the analytical expression for these
quantities in our states (in the case $n=0$) to be
\be \Delta_\pm=\f{W_\pm}{V_\pm^2}-1 =e^{\alpha
^{2}/\sigma^2} \f{\left(1 +\f{3\sigma^2}{4k_0^2}+\f{\alpha
^2}{k_0^2} \right)} {\left(1 +\f{\sigma^2 +\alpha ^2}{4k_0^2}
\right)^2} -1\, .  \label{delta-PM}
 \ee
It is interesting to note that, in the previous expression, there
is some competition between the factor $e^{\alpha ^{2}/\sigma^2}$
and the second part. The exponential becomes close to one if $\sigma$
is large compared to $\alpha$, but can become very large if
$\sigma$ becomes too small. On the other hand, the larger $\sigma$
becomes, the larger the second term. Thus,
that there must be an optimal value $\t\sigma$ for which
$\Delta_\pm$ is the smallest. The form of $\Delta_\pm$ can be
seen in Fig.~\ref{fig:1.1}. Another important aspect to consider is the following.
So far, we have not imposed any condition on $p_\phi$. That is, we have not said that
it has to be `large', mainly because in the classical theory there is no dimension-full
quantity with respect to which one can compare it. In the quantum theory, with the introduction 
of $\hbar$ we do have such scale. As we can see from Eq.(\ref{delta-PM})  and the expression for
$\langle \hat{p}_\phi\rangle$, the quantity $\hbar\alpha$ has the same dimensions of $p_\phi$.
Furthermore, if we now impose the condition $k_0\gg \alpha$, all the terms in parenthesis
in Eq.(\ref{delta-PM}) are small, so we can indeed try to find the optimum value for
$\sigma$. Note also that the condition $k_0\gg \alpha$ implies 
$\langle \hat{p}_\phi\rangle\gg\hbar\alpha$. It is in this sense that we can call
the momentum $p_\phi$ `large'.

The condition that the expression (\ref{delta-PM}) be
an extrema is a cubical equation, and the only physically
interesting solution can be approximated as,
\be \t\sigma^2 \approx 2\,\alpha\, k_0 +\f{2}{7}\, \alpha^2 \, .\ee
%
Since we require that $k_0\gg \alpha$, then the second
term is very small compared to the first one and we can approximate
$\t\sigma\approx \sqrt{2\,\alpha\,k_0}$. It is important to stress that
this condition selects a preferred value for $\sigma$ for which the
state is most semiclassical. As can be seen from the Figure~\ref{fig:1.1},
if we make $\sigma$ slightly smaller (in an attempt to `make the dispersion small')
the asymptotic relative dispersion in volume becomes very large, making the state
not a good candidate for a semiclassical state. 

If we now introduce this value for  $\t\sigma$ into the relative
fluctuation function, then this can be approximated by
  \be
  \left.\lim_{\phi\to\pm\infty}\left[\f{(\Delta \hat{V})^2}{\langle \h V
  \rangle^2}\right]\right|_{\t\sigma}(k_0)
  \approx \f{\alpha}{k_0} \label{opt-dispersion}
  \ee
This approximation has an error of $0.0002 \%$ when $k_0=30000$ (in Planck units), 
and becomes better as $k_0$ grows.
This tells us the precise relation between the parameter $k_0$
that controls the expectation value of $p_\phi$ and the minimum
value of the asymptotic relative fluctuation of the volume
operator. The $\sigma$ dependence of the asymptotic relative
dispersion of the volume
 can be seen  in Figure~\ref{fig:1.1}, with $k_0=30000$.  Two features characterize this
plot, namely, the first one is that for $\sigma^2<\t\sigma^2$ the
function is exponential so it grows very quickly as one approaches zero; the second feature is that for
$\sigma^2>\t\sigma^2$ the function has a polynomial behavior, so one can have values of $\sigma>\tilde{\sigma}$
without $\Delta_\pm$ changing too much.

To summarize, we have seen that not any Gaussian state will be an
admissible semiclassical state. In order to minimize 
the asymptotic relative fluctuation $\Delta_\pm$, one has to choose 
the parameter
$\sigma$ carefully as a function of the other relevant
parameter (in this case, the parameter $k_0$). That is, if we
choose $\sigma$ to be of the order of $\sigma^2 \sim 2\alpha k_0$,
we can then minimize the asymptotic relative dispersion of volume
after and before the bounce (which are the same, as we have seen).
Gaussian states for which this condition is not satisfied will have,
in general, large quantum fluctuations in the regime where we expect them to
be small --for large times after and before the bounce-- so they 
can  not be regarded as semiclassical.

Let us then assume that we have chosen the value $\tilde{\sigma}$ for $\sigma$.
The question we want to ask now is the following: Can we 
reproduce some of the features that are characteristic of the so
called ``effective dynamics''? As we have discussed before, the `classical'  
Hamiltonian constraint one obtains from replacing
classical quantities like connection and curvature by a
corresponding holonomy functions with a finite parameter
$\lambda$, can be seen as either a starting point for `polymer
quantization', or as the `classical limit' of a loop quantized
theory. In any case, we expect that the semiclassical states of
the quantum theory approach this `classical theory' in the
appropriate regime. In other words, we want to know when the
effective dynamics is a good approximation to the full quantum 
dynamics. Is there a choice of parameters in the Gaussian states
for which the effective dynamics is not a good approximation?
We shall answer these questions in the remainder of this section.

\subsection{How Semiclassical is the Bounce?}

\begin{figure}[tbh!]
\includegraphics[scale=0.65]{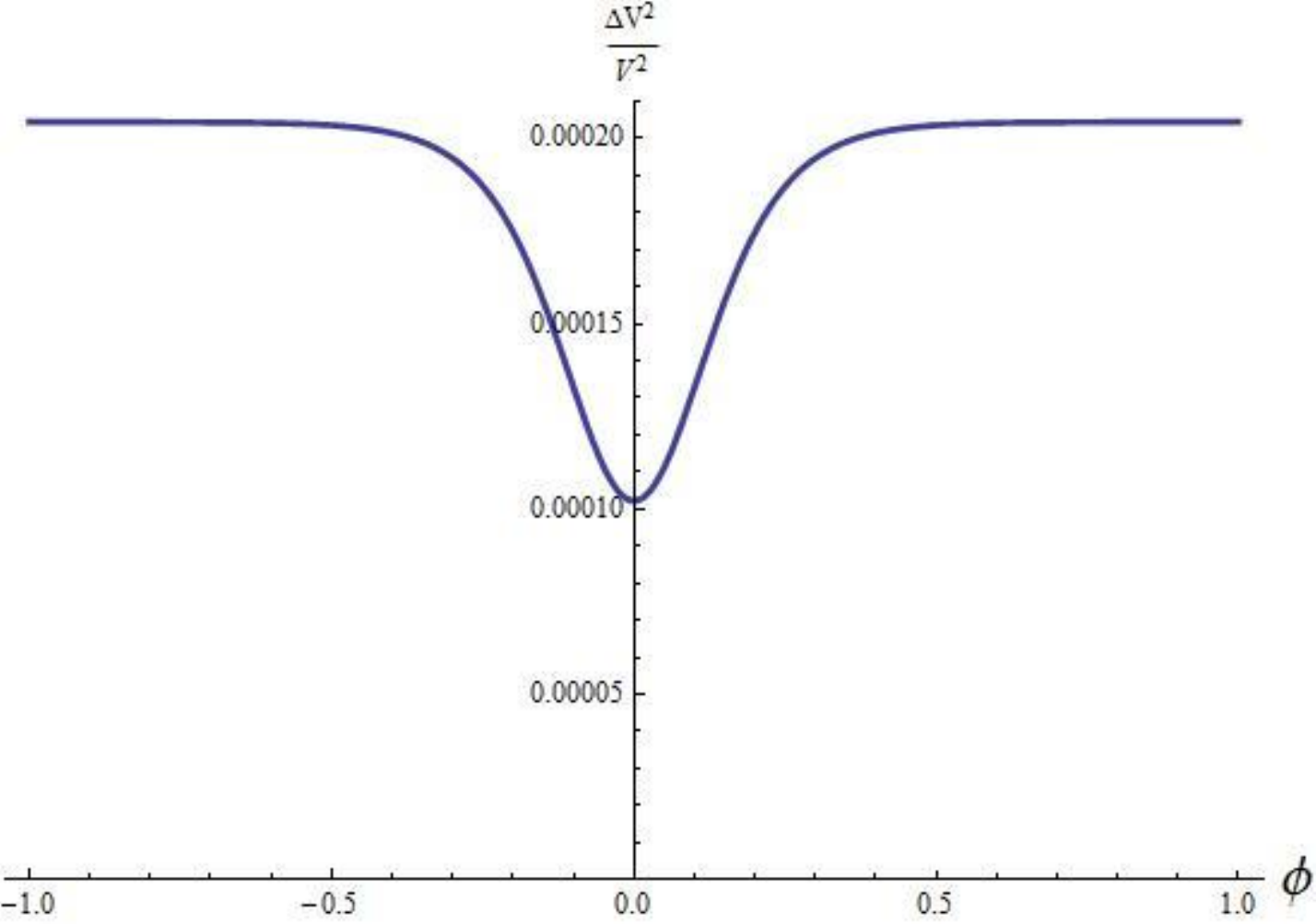}
\caption{The relative fluctuation $(\Delta\h V)^2/\la\hat{V}\ra^2$
is plotted as a function
        of internal time $\phi$ with $k_0=30000$, in Planck units where
        $\alpha\approx 6.14$. Note that the
        minimum of this quantity is at the bounce (here
        corresponding to $\phi=0$). It is found that the value at
        the bounce is approximately 1/2 of its asymptotic value for
        the generalized Gaussian states.
}\label{fig:1.2}
\end{figure}

Another important issue is the question of how semiclassical the
state is at the bounce. For instance, if the state has a very
small relative asymptotic dispersion $\Delta_\pm$ at late times,
one might want to know what the relative dispersion is at the
bounce and how it compares to $\Delta_\pm$. A plot of the relative 
fluctuation, as a function of internal time $\phi$ can be seen in
Fig.~\ref{fig:1.2}, where it is shown that the relative fluctuation
is bounded, it attains its minimum value at
the bounce and approaches $\Delta_\pm$ asymptotically. Thus, somewhat
surprisingly, the state does not become very ``quantum-like"
at the Planck scale but rather preserves its coherence across 
the bounce.

Let us now try to estimate the value of the relative dispersion in
comparison to its asymptotic value. We know that, at the bounce,
$\phib=\f{1}{2\alpha}\ln\left|\f{V_-}{V_+}\right|=\f{1}{4\alpha}\ln\left|\f{W_-}{W_+}\right|$,
then $\la \hat V \ra_{\phib}=  2 \sqrt{V_+V_-}$, and
\be (\Delta \h V )_{\phib}^2 = W_0 -4V_+V_- + 2\sqrt{{W_-}{W_+}}\, .
\ee %
Then the relative fluctuation in the bounce is
\be \left. \f{(\Delta\h V)^2}{\langle\h V \rangle^2} \right|_{\phib} =
\f{W_0 + 2\sqrt{{W_-}{W_+}}}{4V_+V_-}-1\, . \ee
For $n=0$ in the Gaussian states we have 
\be \left.  \f{(\Delta\h V)^2}{\langle\h V \rangle^2}
\right|_{\phib}-\f{1}{2}\left. \f{(\Delta\h V)^2}{\langle\h V
\rangle^2} \right|_{\phinfty} = \f{
e^{-\alpha^2/\sigma^2}\left(1+\f{3\sigma^2}{4k_0^2 }\right)}
{2\left(1+\f{\sigma^2 +\alpha^2}{4k_0^2} \right)^2}-\f{1}{2}\, . \ee
Now, if we impose  $k_0\gg \sigma\gg\alpha$, which is
satisfied if we choose $\sigma=\tilde{\sigma}=\sqrt{2\alpha k_0}$,
then the last equation has the form,
\bas 
\left. \f{(\Delta\h V)^2}{\langle\h V
\rangle^2} \right|_{\phib}-\f{1}{2}\left. \f{(\Delta\h V)^2}{\langle\h
V \rangle^2} \right|_{\phinfty}& = & O(\alpha^2/\sigma^2) +
O(\sigma^2/k_0^2)\, . 
\eas
This tells us that when $k_0\gg \sigma\gg \alpha$, the relative
volume fluctuation at the bounce becomes
 \be \label{eq110} \left.
\f{(\Delta\h V)^2}{\langle\h V \rangle^2} \right|_{\phib}\approx
\f{1}{2}\left. \f{(\Delta\h V)^2}{\langle\h V \rangle^2}
\right|_{\phinfty}=\frac{1}{2} \Delta_\pm\, .
\ee
This approximation, for instance, has an error of $0.007 \%$ when
$k_0=30000$, and it becomes smaller as $k_0$ grows.
%
If we use the Eqs.~\eqref{eq110} and \eqref{opt-dispersion} and choose
the value of $\sigma$ that minimized the relative fluctuation we get 
$$ \left.
\f{(\Delta\h V)^2}{\langle\h V \rangle^2} \right|_{\phib}\approx
\f{\alpha}{2k_0} \, .$$
This approximation have an error of $0.01 \%$
when $k_0=30000$, and becomes smaller as we increase $k_0$. We have performed intensive numerical explorations and have seen that, for $n\geq 1$, the
relative fluctuation is again smaller at the bounce and approaches
$1/2$ of its asymptotic value.

Thus, we can conclude that states that satisfy the
conditions $k_0\gg \sigma\gg \alpha$, 
imposed by our requirements on the asymptotic relative
dispersion of the volume through the choice 
$\sigma=\t\sigma$, exhibit also  semiclassical behavior at
the bounce, in the sense that the relative dispersion in volume is
bounded and small. 
Even more, the relative fluctuation at the bounce is
approximately $1/2$ of the asymptotic value. These analytical results
give support to the numerical explorations reported in \cite{aps,aps2}.

\subsection{Volume and Energy Density at the Bounce}

Having shown that a semiclassical quantum state behaves also
semiclassically at the bounce, it is then natural to ask whether the
state yields observables --through their expectation values-- that
follow some classical trajectories. It turns out that it does, but
the `classical theory' it approaches is not the standard classical
theory (GR+ massless scalar field), but rather an {\it effective theory},
as defined by an effective Hamiltonian constraint ${\cal C}_\l$
\cite{vt}, which contains the {\it quantum geometry scale} $\l$.
The effective dynamics, is generated by the effective
Hamiltonian constraint
\[
{\cal C}_\lambda=-\f{3}{8\pi
G\gamma^2\lambda^2}V^2\sin^2(\lambda\beta)+ \f{p_\phi^2}{2}\approx
0
\]
and has several important features. The first one is that all
trajectories have a bounce that occurs at the same critical
density $\rho_{\rm crit}= \f{3}{8\pi G\gamma^2\lambda^2}$. The
volume reaches a minimum value at the bounce that depends on the
observable $p_\phi$: 
\be V_{\rm min}=\sqrt{\f{8\pi G \gamma^2
\lambda^2}{6}}\;p_\phi\, . 
\ee 
In the quantum theory, we have seen
that every physical states has a bounce where the minimum volume
is given by $V_{\rm bounce}=2\sqrt{V_+\,V_-}$. It is then natural
to find the corresponding volume at the bounce for our family of
Gaussian states. It is straightforward to see that the minimal
value of the volume, for the case $n=0$, is
\be \label{Vmin} V_{\rm bounce}=\f{4\pi \gamma \lp^2
\lambda}{\alpha }\;
 e^{\alpha ^2/2\sigma^2}\, k_0
\left[1 +\f{\sigma^2 +\alpha ^2}{4k_0^2} \right]\, . \ee
which is of the form,
\[
V_{\rm bounce}= \f{4\pi \gamma \lp^2\lambda}{\alpha}\,
 k_0 +{O}(\alpha^2/\sigma^2) +
O(\sigma^2/k_0^2) + O(\alpha^2/k_0^2)\, .
\]
If we 
recall our previous result that $\langle \hat p_\phi \rangle =
\hbar\, k_0 + O(\sigma^2/k_0^2)$, and 
assume the semiclassically conditions found in previous sections, 
namely that $\sigma  \gg
\alpha$ and $k_0 \gg \sigma$ (as is the case when $\sigma=\t\sigma= \sim
\sqrt{2\alpha k_0}$ and $k_0\gg \alpha$) in Eq. \eqref{Vmin}, we can compare
the last two equations to see that $V_{\rm bounce} \approx \f{4\pi \gamma \lp^2
\lambda}{\alpha }\f{\langle \hat p_\phi\rangle}{\hbar}$. 
Using the value $\alpha=\sqrt{12\pi G}$, we conclude that \be
 V_{\rm bounce} \approx V_{\rm min}\, ,
\ee 
which is what we wanted to show. Note that the higher the
value of $k_0$, the better the approximation becomes. In other
words, if $k_0$ were not large enough (compared to $\alpha$), not only 
would the asymptotic relative fluctuation of volume be large,
but the effective equations would fail to be a good approximation
to the quantum dynamics. This seems to indicate that quantum
Gaussian states with a higher value of $k_0$ behave more classically,
in contradiction with the classical intuition that tells us that
`rescaling of $k_0$' is physically irrelevant\footnote{Recall that
the $k$=0 FRW model has a rescaling symmetry of the equations of motion, and 
of the underlying spacetime metric given by $(V,\bb,\phi,p_\phi)\to
(\ell V,\bb,\phi,\ell p_\phi)$ for $\ell$ a constant. This can be
understood as coming from the freedom of choosing arbitrary fiducial
cells for the formulation of the theory \cite{cs:unique}.}. 
Thus, we have to conclude that this rescaling symmetry of the classical
theory is broken in the quantum theory. Note however, that even when the exact
symmetry is broken, one might expect to regain it in the limit of
large $k_0$ which, as we have seen, corresponds to $\la\hat{p}_\phi\ra\gg \hbar\alpha$.
For an in-depth discussion regarding this scaling freedom see \cite{CM-letter}.

Another important result for this solvable model is that the
energy density $\rho$ is absolutely bounded by the critical
density $\rho_{\rm crit}$ \cite{slqc}. It is then natural to ask
what the behavior of the energy density at the bounce is, for
semiclassical states. One might imagine that, for instance, the
more semiclassical the state, the higher the density at the
bounce. In this part we shall explore this question by taking
several quantities to measure density. For instance, the simplest
one would be $\t \rho=\f{\la \h p_\phi\ra^2}{2\la \h V\ra^2}$, a
quantity shown to be bounded by $\rho_{\rm crit}$ \cite{slqc}. It
is straightforward to find this quantity for the $n=0$ states, 
\be
\t \rho=\rho_{\rm crit}\;\f{e^{-\alpha^2/\sigma^2}\left(1 +\f{\sigma^2}{4 k_0^2}
\right)^2} { \left(1 +\f{\sigma^2
+\alpha^2}{4k_0^2} \right)^2}
\ee 
Note that  $\t \rho<\rho_{\rm
crit}$ and one approaches it as $k_0\to \infty$, assuming again
that we have chosen $\sigma \sim \sqrt{2\alpha\,k_0}$ and $k_0\gg \alpha$.
One can also see that, if we were to fix $k_0$ and leave $\sigma$ free,  
the density grows and approaches $\rho_{\rm crit}$ as $\sigma$
grows. Thus, asking for the density to approach the critical density is not a very
stringent condition on $\sigma$ as the relative fluctuations in volume was.
Note also that, if $k_0$ was small enough to be of the order of $\alpha$, the density 
at the bounce would be far from the critical density giving yet another indication
that those states are not semiclassical.
The other quantities representing density that one can
build, by taking for instance $\la \h p_\phi^2\ra$ and $\la\h
V^2\ra$, give expressions that have the same qualitative behavior,
approaching $\rho_{\rm crit}$ as $k_0$ grows.

\subsection{Where is the Classical Region?}

\begin{figure}[tbh!]
\begin{center}
\includegraphics[scale=0.65]{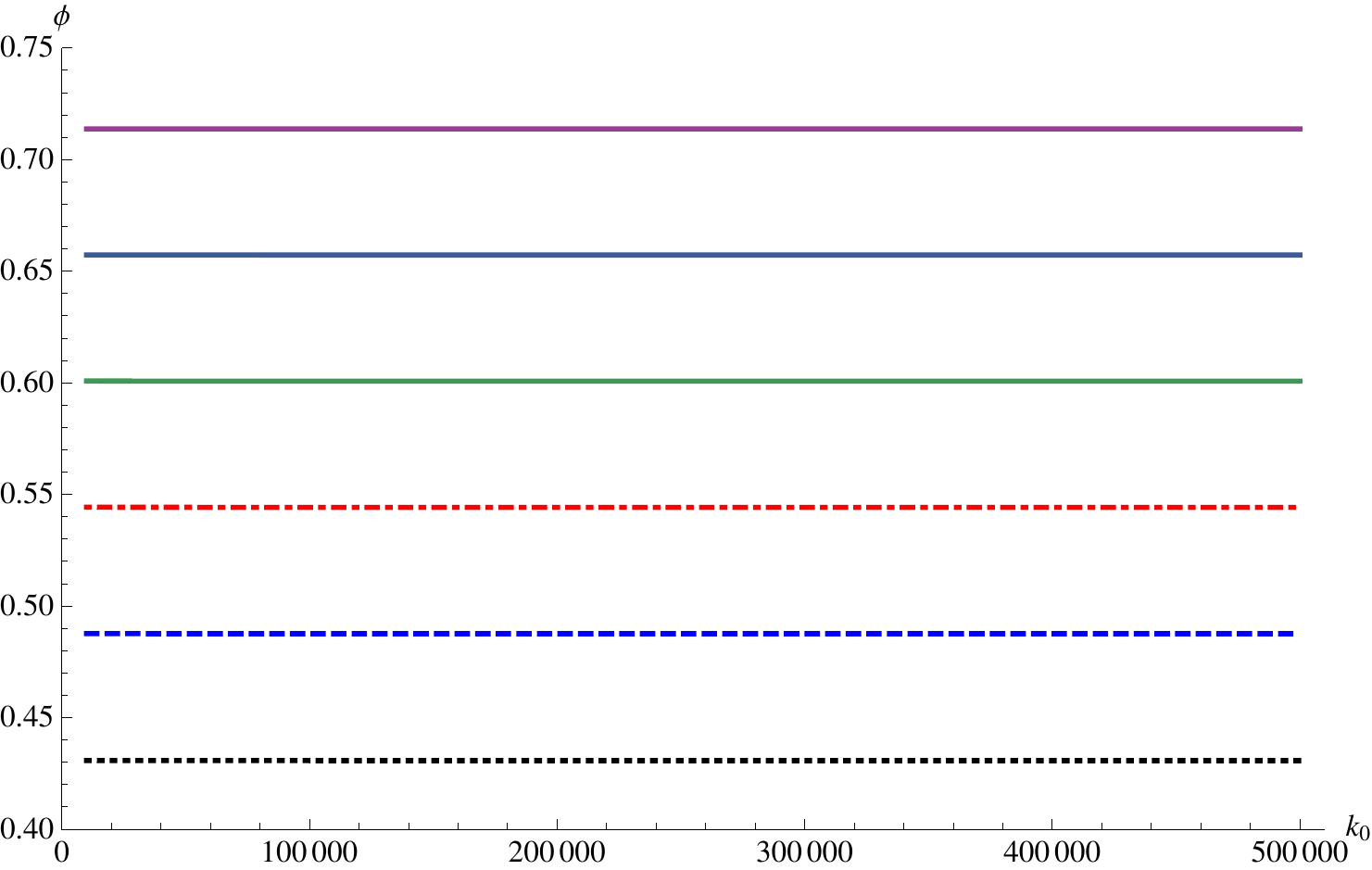}
\caption{The values $\t\phi$ of $\phi$ for which one reaches the classical region are plotted as a function of
$k_0$, where $n$=0, $\sigma=\tilde\sigma$ and $p=0$ at $\phi=0$.
In the plot, $\delta=1$ dotted, $\delta=0.5$ dashed, $\delta=0.25$ dotdashed, $\delta=0.125$ down line, 
$\delta=0.0625$ middle line, and  $\delta=0.03125$ the top line. Note that for high values of $k_0$,
the time $\t\phi$ remains a constant.}
\label{fig:1.3}
\end{center}
\end{figure}
\begin{figure}[tbh!]
\begin{center}
\includegraphics[scale=0.7]{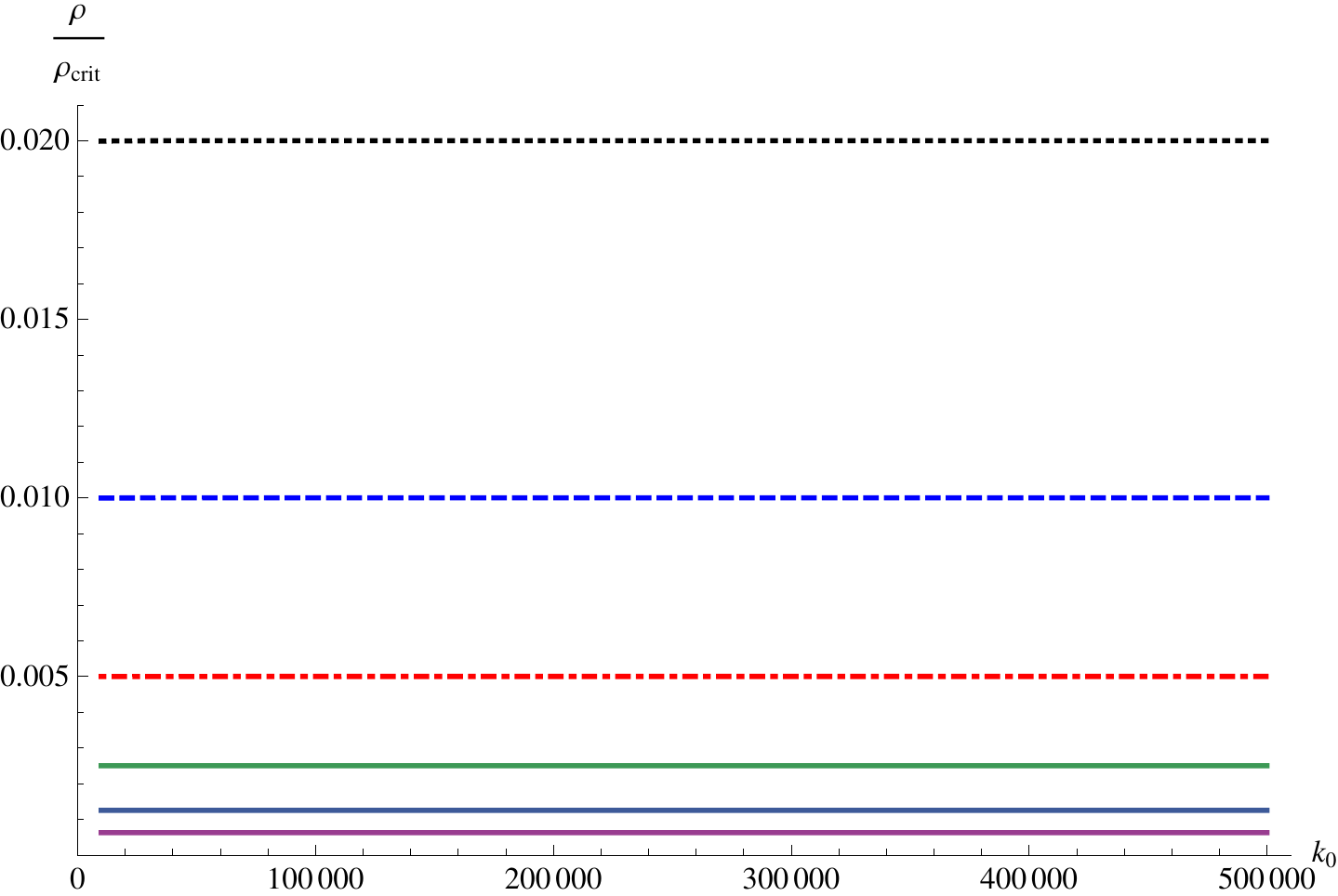}
\caption{Values of density are plotted for different values of $\delta(\t\phi)$, as function of
$k_0$, where $n$=0, $\sigma=\tilde\sigma$ and $p=0$ at $\phi=0$. 
Here, $\rho_{_{1,2,3,4}}/\rho_{\rm crit}$ for which
$\delta=1$ dotted, $\delta=0.5$ dashed, $\delta=0.25$ dotdashed, $\delta=0.125$ up line,
$\delta=0.0625$ middle line, and $\delta=0.03125$ the bottom line.}
\label{fig:1.4}
\end{center}
\end{figure}

In the LQC description of the dynamics as governed by the effective Hamiltonian,
one can unambiguously ask when one recovers the GR classical dynamics. One way of measuring this 
could be in terms of the variable $\bb$ (when $\l\bb\ll 1$), or in terms of the energy density 
($\rho\ll \rho_{\textrm{crit}}$). However if one wants to ask a similar question in the quantum realm, one
immediately faces the following dilemma. As we have seen in the previous part, a Gaussian 
state remains semiclassical across and after the bounce. In fact, the closer to the bounce,
the more semiclassical it is. We can not therefore ask when does the state become semiclassical, based on
the behavior of fluctuations of the relevant operators.
Still, we would like to have the notion of a transition from a quantum gravity dominated regime to 
a the regime where the quantum gravity effects are small and one is in the ``classical regime".

For this purpose, we want to propose such a criteria, and
define the (percentual) error
between the relative fluctuation and the asymptotic relative fluctuation,
\be
\delta (\phi):= \f{\left.\f{(\Delta V)^2}{\la V \ra^2}\right|_\phi -\Delta_+ }{\Delta_+}\times 100 \, .
\ee
We expect that this error will be small when we are approaching the classical region. 
Let us now motivate this proposal. While we can not define a transition in terms of
a change in the relative fluctuations of the basic operators, we could compare them to
those of the standard `Wheeler DeWitt' model. For we know that the expectation values
in the WDW theory follow closely the classical dynamics of general relativity. There
is indeed a canonical way of mapping a LQC state to a WDW state, so the comparison of 
expectation values is well defined \cite{slqc}. We have done that for our Gaussian 
states and computed the corresponding relative fluctuation for the volume. It turns 
out that, for the WDW dynamics, the relative fluctuation is {\it constant} in $\phi$ evolution, and 
corresponds precisely to the asymptotic relative dispersion of LQC. Thus, the LQC evolution
not only approximates the WDW one in terms of expectation values but also in terms of the
relative fluctuations. It is then justified to measure the transition to the `classical
regime' in terms of how close the LQC evolution is to the corresponding WDW dynamics.

In order to
study where the classical region begins, we found the value of $\phi$ for which 
$\delta = 1/2^m$ with $m=0,...,5$, taking $\phi=0$ at the bounce. 
The values $\tilde{\phi}$ of
$\phi$ that satisfy these conditions are plotted in figure \ref{fig:1.3}. Then we
evaluated the densities $\rho_1 = \f{\la \hat p_\phi \ra^2}{2\la \hat V\ra^2}$ , 
$\rho_2 = \f{\la \hat p_\phi \ra^2}{2\la \hat V^2\ra}$ ,
$\rho_3 = \f{\la \hat p_\phi^2 \ra}{2\la \hat V\ra^2}$ ,
$\rho_4 = \f{\la \hat p_\phi^2 \ra}{2\la \hat V^2\ra}$ ,
at this values of $\phi$. The density 
is plotted in figure \ref{fig:1.4}. This plot does not depend
of the used density ($\rho_1 , \rho_2 , \rho_3 , \rho_4$).
From figure \ref{fig:1.4} we can see that
\be
\f{\rho(\tilde{\phi})}{\rho_{\rm crit}}\sim \f{\delta(\tilde{\phi})}{5} \, ,
\label{dens-delta}
\ee
for $k_0\gg \alpha,\, \sigma =\t \sigma = \sqrt{2 \alpha k_0 },\, p=0$ and $n=0$. These results
are similar if $n>0$. 
Furthermore, if we choose $p\ne 0$, the only change is in the value of $\phi$ at the bounce.
If we change the value of $\sigma$ for a fixed $k_0$, then for larger (than $\t \sigma$)
values, the time $\t{\phi}$ of arrival to the classical region is smaller.
In that case, the value of the density is closer to the critical one, and in the limit of large $\sigma$
the classical region is `at the bounce', which does not make much sense. In this way the need  to bound $\sigma$ to be
near the value $\t \sigma$ is manifested.
 
The relation given by Eq.~(\ref{dens-delta}) is telling us the order of the quantum 
corrections. Thus, if we say that the quantum corrections are negligible
when, say, $\delta \sim 10^{-10}$ then the density will be of the order
${\rho}/{\rho_{\rm crit}}\sim 2\times 10^{-11}$.
Recall that Gaussian states are symmetric about the bounce, namely $\Delta_+=\Delta_-$, therefore all the conclusions of this part equally apply before the bounce.

Let us now summarize this section. 
We have seen that appropriate physical conditions imposed on the behavior of basic physical
observables are enough to constraint the  parameters that characterize the generalized Gaussian states (\ref{coh_state}), 
and obtain truly semiclassical states. 
Of the three continuous parameters $(k_0,\sigma,p)$ that characterize the states, two of them are fixed by
semiclassicality considerations, while the choice of $p$ represents a true choice of {\it initial condition}. 
We have seen that asking that the relative fluctuations of the observable $\hat{p}_\phi$ be small imposes that 
$k_0 \gg \sigma$.
Smallness of the asymptotic relative fluctuation of the volume implies then that $k_0\gg\sigma\gg\alpha$.
If these
conditions are satisfied, we saw that the semiclassical states at `late times' remain 
semiclassical across the bounce and that the volume and
density at the bounce are very well approximated by the effective theory. 
We have also seen that the scaling symmetry of the classical theory is not present,
within the class of states under consideration,
when $k_0$ is of the order of $\alpha$. However, in the limit of large
$k_0$, all the properties of the state remain invariant --in terms of being well
approximated by the classical effective theory--. This leads us to conclude that
the scaling symmetry is approximately recovered for ``large $p_\phi$" (for details
regarding this issue, see \cite{CM-letter}).

Still, the Gaussian states we have considered are not the most general states one can consider. In the following section we shall explore the so called {\it squeezed states}, a generalization of the Gaussian states and compare their semiclassicality properties to those of the Gaussian states.

\section{Squeezed States}
\label{sec:5}

In this section we consider states that generalize the Gaussian states
considered so far. The strategy will be to consider squeezed states that are
nearby the Gaussian semiclassical states and compare their properties. In
particular, we would like to know if the squeezed states can improve the
samiclassicality properties such a better behavior of relative fluctuations
of volume or a better approximation to the `effective theory'.

The generalized Gaussian state we have considered so-far had three
free (real) parameters $(p,k_0,\sigma)$ and one discrete parameter
($n$). We saw that, given a point of the physical phase space, we
could approximate the internal dynamics of the theory by a choice
of $p$, and that semiclassicality imposes conditions on the two
other parameters $k_0$ and $\sigma$.

Let us now take the initial states as
\be \label{sq-state}
\t F(k)=\left\{
\begin{array}{rl}
k^n e^{-\eta(k-\beta)^2},& \mbox{ for } k>0 , \,\,\, \eta ,\beta \in {\mathbb C}\\[1ex]
0,& \mbox{ for } k\le 0 ,
\end{array}
\right.
\ee
depending on two complex parameters $(\eta,\beta)$.
One can reduce to a Gaussian state from a squeezed state by setting
\be
\mb{I}{\rm m}(\eta) =:\eta_I =0, \;\;\; \mb{R}{\rm e}(\eta)=: \eta_R=\f{1}{\sigma^2}, \; \;\;
\mb{R}{\rm e}(\beta) =: \beta_R=k_0, \;\;\; 2\eta_R\beta_I=p\, .
\ee
Thus, we see that the extra parameter in the definition of the states if given by the imaginary
part of $\eta$.
The norm of this state in $k$ space  is
\[
(\t\chi,\t\chi)_{\rm phy}=2\int_{-\infty}^\infty k k^{2n}e^{-\eta(k-\beta)^2}e^{-\bar\eta(k-\bar\beta)^2} \d k \, .
\]
Using relation \eqref{expantion} for $t=0$ the last integral can be written as
\be
\|\t\chi\|_{\rm phy}^2=2e^{2\eta_R(a^2-\beta_R^2+\beta_I^2)}e^{4\eta_I\beta_I\beta_R}
\int_{-\infty}^\infty k^{2n+1}e^{-2\eta_R(k-a)^2} \d k\, ,
\ee
where
\be
a = \beta_R-\beta_I\f{\eta_I}{\eta_R} \, .
\ee
%
If we want the  convergence of the integral then $\eta_R>0$ and if we want to select
the positive frequency then $a \gg \f{1}{\sqrt{\eta_R}}$ i.e.
$\beta_R-\beta_I\f{\eta_I}{\eta_R} \gg \f{1}{\sqrt{\eta_R}}$. With this assumptions
the integral is well defined, and the error is small, as shown in Appendix~\ref{app:b}.

If we take $n=0$ and use the change of variables $u=k-a$ we have
\be
\|\t\chi\|_{\rm phy}^2=2e^{2\eta_R(a^2-\beta_R^2+\beta_I^2)}e^{4\eta_I\beta_I\beta_R}
\int_{-\infty}^\infty (u+a)e^{-2\eta_Ru^2} \d k \, .
\ee
Then
\be \label{norm}
\|\t\chi\|_{\rm phy}^2=a\sqrt{\f{2\pi}{\eta_R}}\; e^{2\eta_R(a^2-\beta_R^2+\beta_I^2)}e^{4\eta_I\beta_I\beta_R}\, .
\ee

\subsection{Elementary Observables}

We can compute the expectation value and dispersion of the
fundamental observable $\h p_\phi$. 
\be \langle \hat p_\phi
\rangle= \f{2\hbar}{\|\t\chi\|_{\rm phy}^2} \int_{-\infty}^\infty k^{2n+2}
e^{-\bar\eta(k-\bar\beta)^2}e^{-\eta(k-\beta)^2} \d k . \ee
If we use Eq. \eqref{expantion}, with $t=0$, then we obtain
\be
\langle \hat p_\phi \rangle= \f{2\hbar}{\|\t\chi\|_{\rm phy}^2} e^{2\eta_R(a^2-\beta_R^2+\beta_I^2)}e^{4\eta_I\beta_I\beta_R}
\int_{-\infty}^\infty k^{2n+2}e^{-2\eta_R(k-a)^2} \d k\, .
\ee
If we take $n=0$,
\be
\langle \hat p_\phi \rangle = 
\f{2\hbar}{\|\t\chi\|_{\rm phy}^2} e^{2\eta_R(a^2-\beta_R^2+\beta_I^2)}e^{4\eta_I\beta_I\beta_R}
\left[\f{\sqrt{\pi}}{2}(2\eta_R)^{-3/2}+a^2\sqrt{\f{\pi}{2\eta_R}}\right]\, .
\ee
This becomes, using Eq.~(\ref{norm}),
\be
\la \hat p_\phi \ra =\f{\hbar}{a}\left[\f{1}{4\eta_R} +a^2 \right]\, .
\ee
%
%
Now we want to calculate the expectation value of $\hat p_\phi^2 $.
\be \label{p2-int-sq}
\langle \hat  p_\phi^2 \rangle= \f{2\hbar^2}{\|\t\chi\|_{\rm phy}^2} \int_{-\infty}^\infty k^3 |\t F(k)|^2 \d k .
\ee
In the squeezed states \eqref{sq-state} we get
\be
\langle \hat  p_\phi^2 \rangle= 
\f{2\hbar^2}{\|\t\chi\|_{\rm phy}^2}
 \int_{-\infty}^\infty k^{2n+3} e^{-\bar\eta(k-\bar\beta)^2}e^{-\eta(k-\beta)^2} \d k .
\ee
If we now use Eq. (\ref{norm}), the result is for $n=0$ 
\be
\la \hat p_\phi^2 \ra =\hbar^2 \left[\f{3}{4\eta_R} +a^2 \right] \, .\label{p2-sq}
\ee
From this we can find the dispersion and get,
\be 
(\Delta \hat p_\phi)^2= \f{\hbar^2}{4\eta_R} -\f{\hbar^2}{16 \eta_R^2 a^2}\, . 
\ee
Writing this explicitly in the original variables is
\be
(\Delta \hat p_\phi)^2= \f{\hbar^2}{4\eta_R} -\f{\hbar^2}{16\eta_R^2}\left(\beta_R-\beta_I\f{\eta_I}{\eta_R} \right)^{-2} \, .
\ee

\subsection{Volume}

We shall compute the quantities $V_\pm$ and $W_{0,\pm}$ from which one can find the
expectation value and dispersion of the volume operator.
The relevant coefficients are
\ba \t V_\pm&=&\frac{\kappa}{\alpha} \int_{-\infty}^\infty (-i k)
k^ne^{-\bar\eta(k-\bar\beta)^2} \ i [k\mp i\alpha]
[k\mp i\alpha]^n e^{-\eta(k\mp i\alpha-\beta)^2} \d k \nonumber \\
&=&\frac{\kappa}{\alpha} \int_{-\infty}^\infty k^{n+1}[k\mp i\alpha]^{n+1}e^{-\bar\eta(k-\bar\beta)^2}
 e^{-\eta(k\mp i\alpha-\beta)^2} \d k .
\ea
with ${\kappa}={4\pi \gamma \lp^2 \lambda}$. As in previous
subsections we will only write down explicitly the $n=0$ case.
This expression can be put on the form, 
\be \label{volume} \t V_\pm
=\f{\kappa}{a\alpha}\;\exp\left[ \left(\f{\alpha^2\eta_R}{2} \pm
2\alpha\beta_I\eta_R\right) \left(1+\f{\eta_I^2}{\eta_R^2} \right)
\right] \left[ \f{1}{8\eta_R}+\f{b_\pm^2}{2}
+\f{\alpha^2}{8}\right] \ee
where $b_\pm = a\mp \f{\alpha}{2}\f{\eta_I}{\eta_R}
=\beta_R-\f{\eta_I}{\eta_R} \left( \beta_I \pm\f{\alpha}{2}\right)$.
In the Gaussian variables it takes the form
\ba
\t V_\pm &=&\frac{\kappa}{\alpha}\left(k_0-\eta_I\f{p\sigma^4}{2} \right)^{-1} \;
\exp\left[ \left(\f{\alpha^2}{2\sigma^2} \pm p\alpha \right)
\left(1+\eta_I^2\sigma^4 \right) \right]\times  \nonumber \\
& &\left[\f{\sigma^2}{8}
+\f{1}{2}\left(k_0-\eta_I\sigma^2 \left( \f{p\sigma^2}{2} \pm\f{\alpha}{2}\right)\right) ^2
 +\f{\alpha^2}{8}\right]
\ea
which reduces to the Gaussian case when $\eta_I=0$.

The next quantities we need to compute in order to find the dispersion of the volume
operator are the coefficients $(W_0,W_\pm)$. The first one is given by
\be
\t W_0=2\alpha_2\int_{-\infty}^\infty k^3|\t F(k)|^2 \d k
,\ \mbox{with }\ \alpha_2=\f{2\pi \gamma^2 \lp^4 \lambda^2}{3 G} .
\ee
This is the same integral that the one in Eq. \eqref{p2-int-sq}. Then, in order to 
consider the proper normalization for the squeezed state with  $n=0$, we only
need to replace $\hbar^2$ by $\alpha_2$ into Eq. \eqref{p2-sq}, the result is 
\be
W_0
=\alpha_2\left[\f{3}{4\eta_R} +\left(\beta_R-\beta_I\f{\eta_I}{\eta_R}\right)^2\right] \, .
\ee
The other quantities are given by the expression,
\ba
\t W_\pm
&=&\alpha_3\int_{-\infty}^\infty k^2
\left[\bar{\t F}(k)(k\mp 2i\alpha) \t F(k\mp 2i\alpha) + {\rm C.C.}\right] \d k \, , \nonumber \\
&=&2\alpha_3\mb{R}{\rm e}\left[I_\pm\right]\, ,
\ea
with  $\alpha_3 = \f{\pi \gamma^2 \lp^4 \lambda^2}{3 G}$  and
$I_\pm=\int_{-\infty}^\infty k^2\bar{\t F}(k)(k\mp 2i\alpha) \t F(k\mp 2i\alpha) \d k$.
This integral can be performed for $n=0$ and we get
\be
\t W_\pm =\alpha_3\; \exp\left[ \left(2\alpha^2\eta_R \pm 4\alpha\eta_R\beta_I \right)
\left(1+\f{\eta_I^2}{\eta_R^2} \right) \right] \times
\left[\f{b'_\pm}{2\eta_R a}+\f{1}{4\eta_R}+b_\pm^{'2} +\alpha^2\right]\, ,
\ee
where $ b'_\pm = a\mp {\alpha}\f{\eta_I}{\eta_R}
=\beta_R-\f{\eta_I}{\eta_R} \left( \beta_I \pm \alpha\right)$.
It is now straightforward to find $\Delta_\pm$, the asymptotic relative dispersion for
volume,
\ba
\Delta_\pm
&=&\f{a^2}{4} \;
\exp\left[ \alpha^2\eta_R
\left(1+\f{\eta_I^2}{\eta_R^2} \right) \right]
\f{ \left[\f{b'_\pm}{2\eta_R a}+\f{1}{4\eta_R}+b_\pm^{'2} +\alpha^2\right] }
{\left[ \f{1}{8\eta_R}+\f{b_\pm^2}{2} +\f{\alpha^2}{8}\right]^2}-1', .
\ea
Let us now explore those conditions that select semiclassical states.

\subsection{Semiclassicality Conditions}

In this part, we want to see how the semiclassicality conditions are satisfied if we move
along the `squeezing parameter' $\eta_I$, off the Gaussian semiclassical states, corresponding to
$\eta_I=0$. Instead of performing an exhaustive analysis of this question, as we have done in previous sections for Gaussian states, we have computed the relevant quantities and plotted them.

\begin{figure}[tbh!]
\includegraphics[scale=0.45]{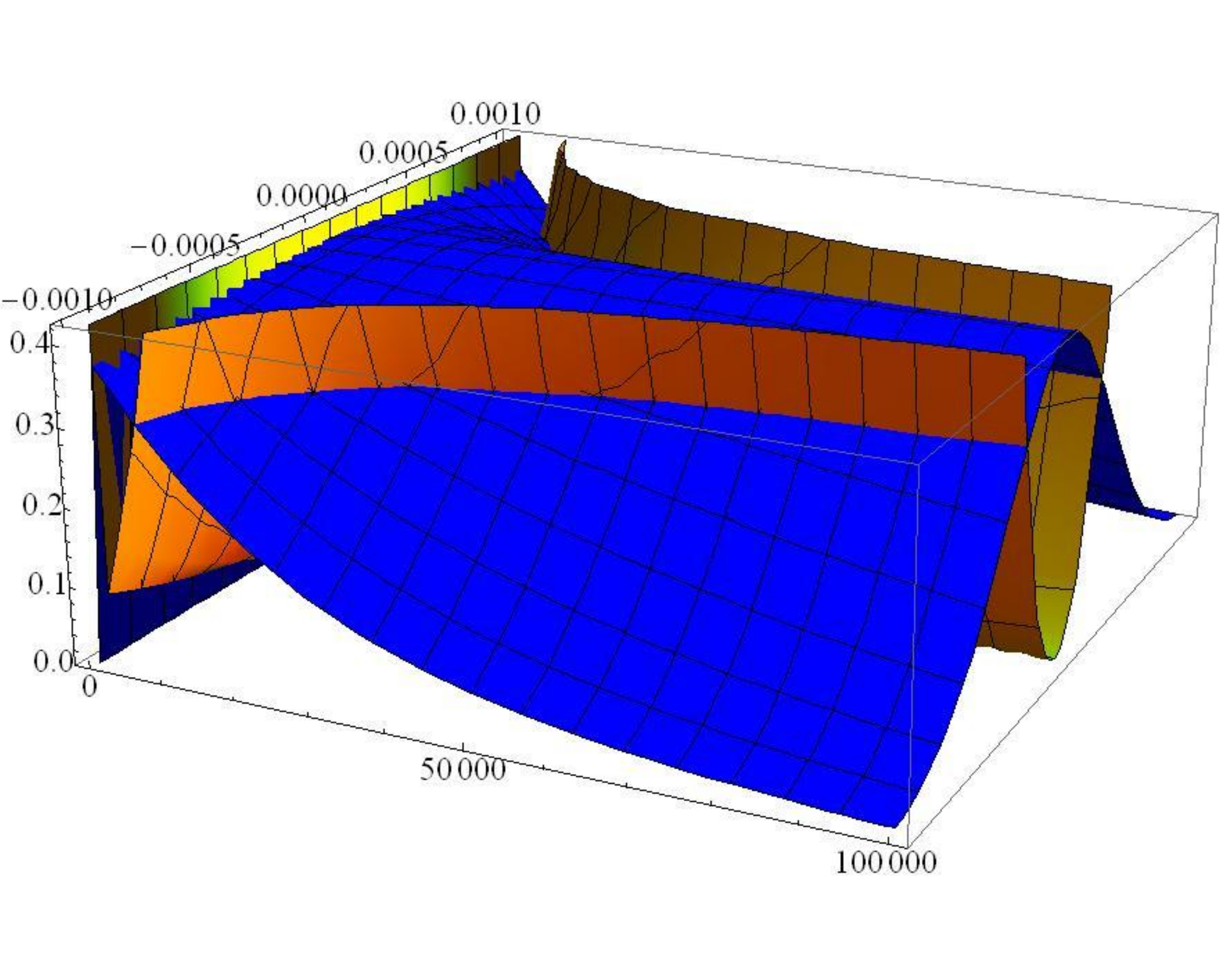}
\caption{Density and asymptotic relative fluctuation $\Delta_+$.
		In the plot $1/\eta_R=\sigma^2: (0, 100 000)$ ,
		$\eta_I:(-0.001,0.001)$ , $\beta_R=k_0=30000$ , $\beta_I=0$ and $n=0$.
        The maximal value of the density correspond to the Gaussian states. 
        Is clear that the squeezed states have a maximal
        density close to $\rho_{\rm crit}$ only if the state is near to the Gaussian states.
}\label{fig:3}
\end{figure}

\vskip0.4cm

Recall that for Gaussian states, asking that the asymptotic relative dispersion in volume be a minimum, selected
an optimal `width' $\sigma=\t \sigma\approx \sqrt{2\alpha k_0}$ for the Gaussian. This value had also the property
that the density at the bounce was very close to the critical density $\rho_{\textrm{crit}}$. In Fig.~\ref{fig:3}
we have plotted  the asymptotic relative dispersion and density at the bounce as functions of $\sigma$ and $\eta_I$.
The $\eta_I=0$ slice corresponds then to the values found for Gaussian states.  It can not be appreciated in the figure but, for a fixed $\sigma$, the asymptotic relative
dispersion $\Delta_\pm$, as a function of $\eta_I$ does not attain its minimum at zero, but rather at some value $\t\eta_I\neq 0$ for $\Delta_+$ (and $\eta_I=-\t\eta_I$ for $\Delta_-$). Thus, if one chooses, for the parameter of
the squeezed state the value $\eta_I=\t\eta_I$, one is `improving' the asymptotic relative dispersion after the bounce,
and, at the same time, increasing the value of the asymptotic dispersion {\it before} the bounce. 
In this process, one is introducing an asymmetry
in the fluctuations. This fact has been used to suggest the possibility, for example, that the asymmetry 
could be so large to
spoil the semiclassicality properties across the bounce. As we shall see below, this possibility is however, not realized. 
What one can indeed see from the figure is that,
as we move further away from the Gaussian states, that is, as we increase to values $|\eta_I|>|\t\eta_I|$, the asymptotic relative dispersion increases steeply, for both signs of $\eta_I$.
As can be seen from the figure, the density  at the bounce (in blue) decreases as we go away from
$\eta_I=0$ in both directions, reaching densities much less than the critical value very fast.

\begin{figure}[tbh!]
\includegraphics[scale=0.4]{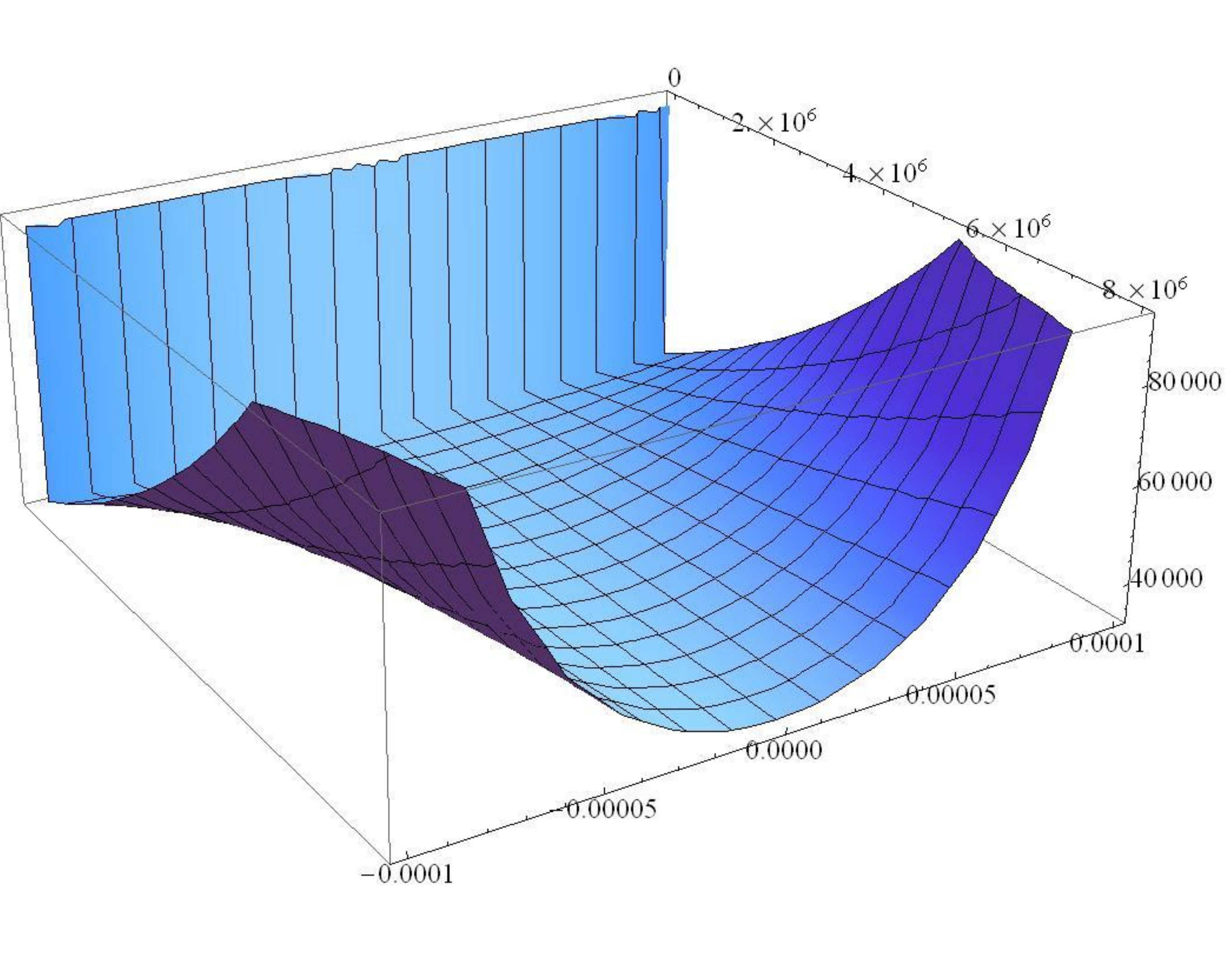}
\caption{Volume at the Bounce. 
		In the plot $1/\eta_R=\sigma^2: (0, 8\times 10^6)$ ,
		$\eta_I:(-0.0001,0.0001)$ , $\beta_R=k_0=30000$ , $\beta_I=0$ and $n=0$.
		The global minimum
        is over the Gaussian states and has an exponential behavior
        if we move far away from the Gaussian state or if $\sigma^2$
        is smaller that $\t\sigma^2=2\alpha k_0$.
}\label{fig:2}
\end{figure}

Both phenomena suggest that we can indeed consider squeezed states as semiclassical states {\it provided we remain
very close to the Gaussian states}. The question is then how close we can be in order to still have semiclassical behavior as exhibited by the Gaussian states. For this purpose, we have plotted in Fig.~\ref{fig:2} the volume at the bounce
as function of $\sigma$ and $\eta_I$. We again see that, as was the case for the density at the bounce, the volume at 
the bounce has indeed a minimum at the Gaussian states, namely, for $\eta_I=0$. Using both relative volume and density
at the bounce, we can then define an interval, as function of $\sigma$, in which the parameter $\eta_I$ can take values and the state still be considered as semiclassical. More precisely, we define some tolerance for the asymptotic relative
dispersion $\Delta_+=\t\Delta$, and find the maximum value of $\eta^{\textrm{max}}_I$ for which the relative dispersion is below the
value $\t\Delta$. We then find the value of the density at the bounce for the value  $\eta^{\textrm{max}}_I$. We have made extensive explorations for different values of $\t\Delta$ and $k_0$, and have found that, for a fixed value of $\t\Delta$, the larger the value of $k_0$, the smaller the allowed interval in $\eta_I$. For example, for $\t\Delta=0.01$, the dependence
of $\eta^{\textrm{max}}_I$ on $k_0$ can be approximated (for large $k_0$, $\sigma=\t\sigma$, $n=0$ and $p=0$) as $\eta^{\textrm{max}}_I\sim 1/\sqrt{47000\,k_0}$. The value of $\rho_{\textrm{b}}$ at that point is equal to $0.99\rho_{\textrm{crit}}$.
(For $\t\Delta = 0.001$, $\eta^{\textrm{max}}_I\sim 1/\sqrt{470000\,k_0}$ and $\rho_{\textrm{b}} = 0.999\rho_{\textrm{crit}}$.)

\begin{figure}[tbh!]
\includegraphics[scale=0.45]{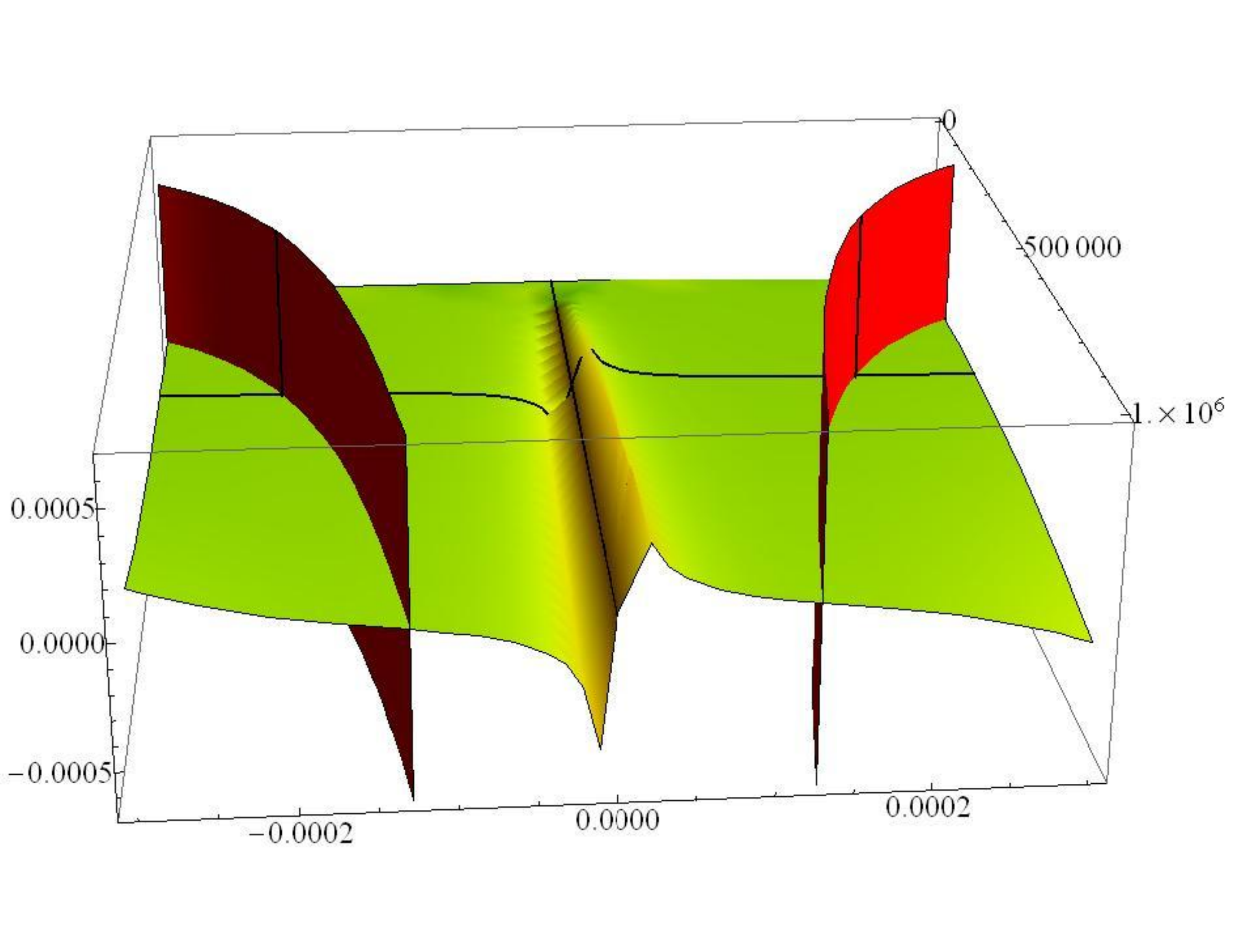}
\caption{Relative error of the difference in the asymptotic values of the
        relative fluctuation $\frac{\Delta_+-\Delta_-}{\Delta_-}$. 
		In the plot $1/\eta_R=\sigma^2: (0, 10^6)$ ,
		$\eta_I:(-0.0003,0.0003)$ , $\beta_R=k_0=30000$ , $\beta_I=0$ and $n=0$.
        Outside of the vertical one
        the squeezed state have an asymptotic relative fluctuations 
        ($\Delta_+$ and $\Delta_-$) bigger that 1.
        One line mark the Gaussian states ($ \eta_I=0$) that is just when the curve is zero 
        and the other line is for states with $\t\sigma^2=2\alpha k_0$.
}\label{fig:4}
\end{figure}

Let us now return to the issue of the asymmetry across the bounce of the relative dispersion in volume. As we have mentioned, without any control on the value of the parameters, one could in principle have a large asymmetry that translates into a loss of semiclassicality across the bounce. This scenario was suggested in \cite{harmonic}, but where the analysis was made for an effective formalism not directly derived from LQC and with little control on the semiclassicality conditions. Here we are better equipped to make precise statements about this asymmetry based on the control we have on analytical expressions and the possible values the parameters of the states can have. In Fig.~\ref{fig:4} we have plotted the relative difference $\Delta_+/\Delta_- - 1$ in asymptotic relative dispersion, as a function of $\sigma$ and
$\eta_I$. We have included, as vertical surfaces, the boundary of the allowed region in parameter space. Two slices
in the figure are worth mentioning. The curve $\eta_I=0$, which is viewed almost vertical in the figure corresponds to the Gaussian states. On that curve, $\Delta_+=\Delta_-$ so the relative difference vanishes. The second highlighted curve corresponds to the choice $\sigma=\t \sigma$, which is the extension to $\eta_I\neq 0$ of the optimal Gaussian states. 
Note that this curve starts at zero, increases (for positive $\eta_I$), reaches a maximum and then decreases tending to
zero. The curve is antisymmetric with respect to $\eta_I$. Thus, there are two points for which $|\Delta_+/\Delta_- - 1|$ 
reaches a maximum. What we have seen, by taking many different values of $k_0$, is that this maximum value is indeed,
extremely small. For instance, for $k_0=10^5$, the maximum is about $10^{-4}$. As we increase $k_0$ this value decreases even further. These results show that indeed, semiclassical states are very symmetric and that semiclassicality is preserved across the bounce. One should also note that this quantitative analysis invalidates the claims of \cite{odc}, and \cite{harmonic} regarding the allowed asymmetry in fluctuations across the bounce and `cosmic forgetfulness', and supports the results of \cite{cs:prl} and \cite{kam:paw}\footnote{One should recall that the model defined in \cite{odc}, even though using a quantization quite different than that in \cite{aps2}
could be seen, in a loose sense, as a different factor ordering from the choice made in LQC and here. For a discussion on the differences between the
assumptions made in the model of \cite{odc} and standard LQC see \cite{slqc}, and for a detailed criticism of some claims made in \cite{odc}, see \cite{cs:prl}. Our results here can be seen as giving further validity to the arguments presented in \cite{cs:prl} against the claims of \cite{odc,harmonic}.}. For instance, in the first reference of \cite{harmonic}, it is claimed that the relative change in relative asymptotic dispersion in volume can be as large as 20 for semiclassical states, while we have demonstrated that this quantity is several orders of magnitude smaller for realistic values of the parameter $k_0$ (of the order of $10^{-4}$ for $k_0=10^5$, and much smaller for the values expected to represent realistic universes ($k_0 \sim 10^{100}$ or larger)).

\section{Discussion}
\label{sec:6}

Let us summarize our results. We have defined coherent Gaussian states as candidates for semiclassical states
`peaked' around points of the classical (physical) phase space. We have imposed consistency conditions on such
states by asking that the relative dispersion of the volume be small at late times after the bounce. This 
condition implies that one of the parameters, yielding the momentum of the scalar field, be large and fixes
one of the other parameters. We can therefore find a canonical Gaussian state for a given point of the 
classical phase space corresponding to a classical solution at late times. When exploring the properties 
of these states near the quantum bounce we found that they behave also semiclassically in the deep quantum region; the relative fluctuations of volume are of the same order and smaller than in the asymptotic, large volume, region. Furthermore, for those states that have small fluctuations, the expectation value of the volume and energy density at the bounce are very
well approximated by the so called effective theory, thus giving support to the claim that this theory is a good approximation to the quantum dynamics, even in the deep Planck regime.
Next, we introduced squeezed states, a one parameter generalization of Gaussian states and studied their properties in a vicinity of the semiclassical Gaussian states. As we showed, the range of this extra parameter is severely restricted if we want to maintain a small relative fluctuation for the volume. 
Furthermore, if one departs from the Gaussian states too much,
both the volume and density at the bounce differ very rapidly from the value on the Gaussian states (and also the `effective theory'). Thus, we are lead to conclude that the Gaussian states exhibit a better semiclassical behavior than squeezed states.

An important issue that can be studied quantitatively in this solvable model is that of the asymmetry on the volume fluctuations. On Gaussian states the fluctuations are the same before and after the bounce. As one introduces the squeezing parameter, the fluctuations are no longer symmetric. In fact, depending on the choice of sign of the parameter, either one of
the fluctuations becomes smaller and reaches an absolute minimum in the vicinity of the Gaussian states. At the same time, the relative fluctuation on the opposite side of the bounce increases. The question is then how big can this asymmetry be? What we saw is that for a given, somewhat arbitrary  choice of parameters, that is still 
far from those giving rise to a
realistic `large' universe, the relative difference in relative dispersion is very small (of the order of $10^{-4}$), and decreases for more realistic values of the parameters.
Thus, even if we were to choose the squeezing parameter such that this asymmetry is maximized, the relative change is so small that, for all practical purposes, the state is symmetric. A state that is semiclassical on one side of the bounce not only remains semiclassical on the other side, but maintains its coherence. This quantitative analysis therefore invalidates claims of `loss of coherence' across the bounce \cite{harmonic}, where it was claimed that the relative change in (relative) dispersion can be several orders of magnitude larger. 

One particular feature of the classical description of the system under consideration is that it possesses a rescaling symmetry. In the spacetime description, we can rescale the volume by a constant and the physical properties of the spacetime remain invariant. 
In the phase space description this is manifested by a symmetry in the equations of motion. This symmetry, however,  can not be treated as {\it gauge}, in the same sense that Hamiltonian symmetries are.
One can understand the origin of this symmetry by recalling that the $k$=0 model needs a fiducial volume for its Hamiltonian description, and this choice is completely arbitrary. In turn, this means that there is no scale with respect to which one could compare the momentum  $p_\phi$ of the scalar field; there is no meaning to the statement that `$p_\phi$ is large'. As we have seen, in the quantum theory Planck constant introduces such a scale  and this is manifested in the behavior of certain quantum observables. One of the semiclassicality conditions we have found is that the larger  the expectation value of $\langle \hat{p}_\phi\ra$, the more semiclassical the state is. What we have also seen is that the rescaling symmetry is only approximately recovered, for our class of
states, in the `very large $\la \hat{p}_\phi\ra$ limit', which can be taken as the {\it classical limit} of the system. For further details on this issue, see \cite{CM-letter}.
 
\section*{Acknowledgments}

\noindent We thank A. Ashtekar, Y. Ma, T. Pawlowski and P. Singh for helpful discussions and comments.
This work was in part supported by DGAPA-UNAM IN103610, by NSF
PHY0854743 and by the Eberly Research Funds of Penn State.

\begin{appendix}

\section{Effective Reduced Phase Space}
\label{app:a}

The kinematical phase space of the model is $\Gamma=\mb{R}^4$ with coordinates
$(\bb,V,\phi,p_\phi)$. Here $\phi$ is the matter scalar field, $p_\phi$ its canonical
momentum, $V$ the volume (of the fiducial cell) and $\bb$ its canonical momentum.
The effective Hamiltonian constraint of the system is \cite{vt}:
\[
H_{\l}=\f{3}{8\pi G\gamma^2 \lambda^2}V^2 \sin(\lambda
\bb)^2-\f{p_\phi^2}{2}\approx 0
\]
which is labeled by a real parameter $\lambda$ with dimensions of
length. The symplectic structure is given by:
\be \label{forma}
 \Omega_0=\f{1}{4\pi G\gamma}\d V\wedge \d\bb+\d
p_\phi\wedge \d\phi
\ee
which induces the following Poisson brackets
\[
\{\bb ,V\}_0=4\pi G\gamma\ \ ,\ \ \{\phi,p_\phi\}_0=1\, .
\]
The Hamiltonian constraint defines a hypersurface $\bar{\Gamma}$
in phase space. It can be taken as a disjoint union of
hypersuperfaces where $\bb \in
(n\f{\pi}{\lambda},(n+1)\f{\pi}{\lambda})$ with $n \in\mb{Z}$. In what
follows we shall only consider the connected component where
$\bb \in (-\f{\pi}{\lambda},0)$. Note that the points where
$\bb=n\f{\pi}{\lambda}$, $n \in\mb{Z}$ are not in the constrained surface.
These are degenerate points (hyperplanes) where the system does not have
a well defined dynamics. We shall see that physically, these points
should be excluded.

We shall now consider the pullback $\bar\Omega$ of the symplectic
structure to the hypersurface $\bar\Gamma$. That is, we find
$\bar\Omega$ such that $\Omega_0=\iota*\bar\Omega$, for
$\iota:\bar\Gamma \to \Gamma$ the embedding of $\bar\Gamma$ into
$\Gamma$.

Let us write
\[
H_\l=\eta^2 V^2 \sin(\lambda \bb)^2-p_\phi^2\approx 0\ , \ \mbox{with}
\ \ \ \eta^2=\f{3}{4\pi G\gamma^2 \lambda^2}
\]
then, on the constrained surface we have
\be \label{p-on-surface}
p_\phi\approx \pm\eta V\,\sin(\lambda \bb)
\ee
The pullback of the gradient of $H_\l$ to $\bar\Gamma$ yields,
\ba \eta^2 2V \d V \sin(\lambda \bb)^2
+2\eta^2 V^2 \sin(\lambda \bb) \cos(\lambda \bb)\lambda \d\bb
\approx 2 p_\phi \d p_\phi\\
\pm \eta[\sin(\lambda b)\d V + V \cos(\lambda \bb)\lambda \d \bb] \approx \d
p_\phi \ea
which can be used for solving $\d p_\phi$,  $\d V$ or  $\d\bb$ and
inserting back in Eq.~(\ref{forma}). Each of this choices will depend
on our choice of coordinates for the constrained surface. Let us study
two of such parametrizations. In the first one we solve for
$p_\phi$, which means that we use $(V,\bb,\phi)$ as coordinates. Then
\[
\bar\Omega_1=\f{1}{4\pi G\gamma}\d V\wedge \d\bb \pm
\eta[\sin(\lambda \bb)\d V + V\,\cos(\lambda \bb)\lambda \d\bb]\wedge \d\phi
\]
which gives us the expression for the two disconnected
hypersuperfaces parametrized by positive and negative values for
$p_\phi$. The value $p_\phi=0$ is excluded if $\bb\ne
n\f{\pi}{\lambda}$ or $V\ne 0$, with  $n \in\mb{Z}$. If we take
$p_\phi >0$ we get
\[
\bar\Omega_1=\f{1}{4\pi G\gamma}\d V\wedge \d\bb + \eta\, \sin(\lambda
\bb)\d V\wedge \d\phi + \eta\,\lambda\, V \cos(\lambda \bb)\, \d\bb\wedge d\phi
\]

If we now solve for $\d V$, we get the expression for the pre-symplectic structure
in the $(\bb,\phi,p_\phi)$ parametrization as,
\[
\bar\Omega_2=\pm\f{1}{4\pi G\gamma}\f{1}{\sin(\lambda
\bb)\eta}\d p_\phi \wedge \d \bb+\d p_\phi\wedge \d\phi
\]
We can again restrict ourselves to $p_\phi >0$, which yields
\[
\bar\Omega_2=\f{1}{4\pi G\gamma \eta}\f{1}{\sin(\lambda
\bb)}\d p_\phi \wedge \d \bb+ \d p_\phi\wedge \d\phi
\]
on the corresponding connected component. Regardless of the
parametrization chosen, the pre-symplectic form $\bar\Omega$ has a
degenerate direction corresponding to the Hamiltonian Vector field
$X_{H_\l}$ of the Hamiltonian constraint $H_\l$. Physically the
integral curves represent {\it gauge} direction along which points
are physically indistinguishable. We can find the gradient of the
constraint,
\[
\nabla_a H=2\eta^2 V \sin(\lambda \bb)^2 \nabla_a V
+2\eta^2\lambda V^2 \sin(\lambda \bb)\cos(\lambda \bb)\nabla_a \bb
- 2 p_\phi \nabla_a p_\phi
\]
and, using that $X^a_H=\Omega^{ab}\nabla_b H$, and
\[
\Omega^{ab}=2\left(4\pi G\gamma \f{\partial^{[a}}{\partial
\bb}\f{\partial^{b]}}{\partial V}
+\f{\partial^{[a}}{\partial\phi}\f{\partial^{b]}}{\partial p_\phi}
\right)
\]
we get,
\[
\left. X^a_H\right|_{\bar\Gamma}= 16\pi G\gamma\eta^2 V
\sin(\lambda \bb)^2\,\left(\f{\partial}{\partial \bb}\right)^a
-16\pi G\gamma\eta^2\lambda\, V^2 \sin(\lambda \bb)\cos(\lambda
\bb)\left( \f{\partial}{\partial V}\right)^a - 4
p_\phi\,\left(\f{\partial}{\partial\phi}\right)^a
\]
as the restriction of the vector field to the constrained
hypersurface with coordinates $(V,\bb ,\phi)$, using Eq.~\eqref{p-on-surface}. 
It is easy to check that this is indeed the degenerate direction of
$\bar\Omega$: $\bar\Omega_1(X_H,\cdot )=0$.

We are now interested in understanding qualitatively the structure
of the reduced phase space, for which we would need to take the
quotient of $\bar\Gamma$ by the gauge directions generated by
$X_H$. In this case, it turns out to be easier to perform a gauge
fixing which, as we shall see, is well defined everywhere. Let us
define ${\cal C}=\phi-c$, with $c=constant$. It is direct to see
that it forms, with $H_\l$ a second class pair,
\[
\{H,{\cal C}\}_0=\{H,\phi\}_0= \{\eta^2 V^2 \sin(\lambda
\bb)^2-p_\phi^2,\phi\}_0= -\{p_\phi^2,\phi\}_0=2p_\phi
\]
provided $p_\phi\ne 0$, which is a condition we had asked before.
The gauge fixed phase space $\hat{\Gamma}$ is therefore parametrized
by $(\bb,p_\phi)$ (or $(V,\bb)$) with $\phi=constant$, $p_\phi>0$
or $p_\phi<0$ and $\bb \in (-\f{\pi}{\lambda},0)$. The
corresponding symplectic structures on each of this two different
parametrizations are
\[
\hat{\Omega}_1=\f{1}{4\pi G\gamma}\;\d V\wedge \d\bb\qquad
,\qquad \hat{\Omega}_2=\pm\f{1}{4\pi G\gamma \eta \sin(\lambda
\bb)}\;\d p_\phi \wedge \d\bb
\]
If we introduce the new variable (as was done in the main text),
\[
x=\f{1}{\sqrt{12\pi G}}{\rm ln}\left| \tan\left(\f{\lambda
\bb}{2}\right) \right|
\]
then the symplectic form $\hat{\Omega}_2$ becomes
\[
\hat{\Omega}_2=\pm \d p_\phi \wedge \d x
\]
which indicates that $(x,p_\phi)$ can be seen as conjugate variables
in the reduced theory.


\section{Some Integrals}
\label{app:b1}

Here we list some integrals that are useful in the main part of the 
manuscript.
\be
\int_{-\infty}^\infty e^{-\eta x^2}\d x=\sqrt{\f{\pi}{\eta}}
\ee
%
%
%
%
\be
\int_{-x_0}^{x_0} x^{2n+1} e^{-x^2}\d x=0, \ \ \ \mbox{with }\ n \in \mb{Z}
\ee
\be
\int_{-\infty}^\infty x^{2n} e^{-\eta x^2}\d x=
(-1)^n\f{\d^n }{\d \eta^n}\sqrt{\f{\pi}{\eta}}
,\ \ \ \mbox{with }\ n \in \mb{Z}
\ee
Some error integrals.
\be
\int_{z}^\infty e^{-x^2}\d x =
\f{\sqrt\pi}{2}erfc(z)
,\ \ \ \mbox{with }\ z \in \mb{C}
\ee
\be
\int_{z}^\infty x^2e^{-x^2}\d x =
\f{z}{2}e^{-z^2}+\f{\sqrt\pi}{4}erfc(z)
,\ \ \ \mbox{with }\ z \in \mb{C}
\ee
where $erfc(z)$ is the Complementary Error Function
\be
erfc(z)=1-erf(z)=1-\f{2}{\sqrt\pi}\int_{0}^z e^{-x^2}\d x
\ee
%
which can be bounded using its asymptotic expansion \cite{abramowitz} as
\be\label{erfc_approx_complex}
|erfc(z)|<\f{e^{-|z|^2}}{|z|\sqrt\pi}
\ee
with $|$arg $z|<\pi/4$ and $z\rightarrow\infty$. When $z=x$ is real then
\be\label{erfc_approx}
erfc(x)<\f{e^{-x^2}}{x\sqrt\pi}
\ee

\subsection{Some Relations}
\vspace{-1cm}
\ba
-\bar\eta(k-\bar\beta)^2-\eta(k\mp i t -\beta)^2 &=& -2\eta_R\left(k-a\mp \f{i\eta t}{2\eta_R}\right)^2 \notag \\
& &+\left(\f{t^2\eta_R}{2}\mp 2t\beta_I\eta_R\right)\left(1+\f{\eta_I^2}{\eta_R^2} \right) \notag \\
& &+2\eta_R(a^2-\beta_R^2+\beta_I^2)+4\eta_I\beta_I\beta_R \label{expantion}
\ea
\be \label{product}
\left[u+a\pm \f{i\eta \alpha}{2\eta_R}\right]\left[u+a\pm \f{i\eta \alpha}{2\eta_R}\mp i \alpha\right]
=u^2+2u b_\pm+b_\pm^2+\f{\alpha^2}{4}
\ee

\section{Errors}
\label{app:b}

We show in explicit form the errors for the norm and volume and
explain why the approximations we take throughout the manuscript are completely justified.

\subsection{Norm Error}
The norm of the Gaussian states is given by
\ba \label{norm_k}
\|\t\chi\|_{\rm phy}^2&=&2\int_{0}^\infty k k^{2n}e^{-2(k-k_0)^2 /\sigma^2} \d k \\
\label{norm_int2}
&=&2\left[\f{\sigma}{\sqrt{2}}\right]^{2n+2}
\int_{-\sqrt{2}k_0/\sigma}^\infty \left[u+\f{\sqrt{2} k_0}{\sigma}\right]^{2n+1} e^{-u^2} \d u
\ea
which for the $n=0$ case becomes,
\be
\|\t\chi\|_{\rm phy}^2=2\left[\f{\sigma}{\sqrt{2}}\right]^{2}
\int_{-y}^\infty \left[u+y\right] e^{-u^2} \d u\; , \;\;\;
\mbox{with} \;\;\;
y=\f{\sqrt{2} k_0}{\sigma}\, .
\ee
Then
\bas
\|\t\chi\|_{\rm phy}^2
&=&\sigma^2 \int_{-y}^\infty u e^{-u^2} \d u+\sigma^2\int_{-y}^\infty y e^{-u^2} \d u\\
&=&\f{\sigma^2}{2}e^{-y^2}+\sigma^2 y\sqrt\pi
-\sigma^2 y\f{\sqrt\pi}{2}erfc(y)\, .
\eas
Using the form of $y$, we get
\be
\|\t\chi\|_{\rm phy}^2=\sqrt{2\pi} k_0\sigma +\f{\sigma^2}{2}e^{-(\f{\sqrt{2} k_0}{\sigma})^2}
-\sigma k_0\sqrt{\f{\pi}{2}} erfc\left( \f{\sqrt{2} k_0}{\sigma}\right) \, .
\ee
The norm that we used in the manuscript was
\be \label{norm0-ap}
\|\t\chi\|_{\rm phy}^2\approx \sqrt{2\pi} k_0\sigma.
\ee
Then the error in this approximation is just
\be
{\rm Error}(\|\t\chi\|_{\rm phy}^2)=\f{\sigma^2}{2}e^{-(\f{\sqrt{2} k_0}{\sigma})^2}
-\sigma k_0\sqrt{\f{\pi}{2}}erfc\left( \f{\sqrt{2} k_0}{\sigma}\right)\, .
\ee
This tells us that the error in the norm for $n=0$ is the error in the approximation
of the function $erfc(x)$ by the function $\f{e^{-x^2}}{x\sqrt\pi}$.
If we use Eq. \eqref{erfc_approx} to bound the function $erfc(y)$ then
the error can be bounded by
\be
{\rm Error}(\|\t\chi\|_{\rm phy}^2 )
 <  \f{\sigma^2}{2}e^{-(\f{\sqrt{2} k_0}{\sigma})^2}\, .
\\
\ee
Since we have seen in the main text that semiclassicallity
implies that we take $ k_0 \gg \sigma >0$ then this error is indeed very small.
These errors were also studied numerically for $n>0$ and present similar behavior.
We can then conclude that the approximation made is very good when computing the norm of
the states.


\subsection{Volume Error}
The coefficients in the expectation value for volume are,
\ba
\t V_\pm &=&\f{\kappa}{\alpha} e^{\pm
p\alpha}\int_{0}^\infty [k(k\mp i\alpha)]^{n+1}
e^{-(k-k_0)^2 /\sigma^2}e^{-(k\mp i\alpha-k_0)^2 /\sigma^2} \d k \\
\label{vol_int}
&=&\f{\kappa}{\sqrt{2}\alpha}\left(\f{1}{4}\right)^{n+1}
e^{\pm p\alpha} e^{\alpha^2/2\sigma^2}\int_{-u_0}^\infty
\left[2u^2+4\sqrt{2}k_0 u+4k_0^2 +\alpha^2 \right]^{n+1}
e^{-u^2/\sigma^2}\d u\, ,
\ea
with $u_0 = \sqrt{2}k_0\pm i\alpha/\sqrt{2}$. Then for $n=0$ are
\be
\t V_\pm=\f{\kappa}{\sqrt{2}\alpha}\f{1}{4}
e^{\pm p\alpha} e^{\alpha^2/2\sigma^2}\int_{-u_0}^\infty
\left[2u^2+4\sqrt{2}k_0 u+4k_0^2 +\alpha^2 \right]
e^{-u^2/\sigma^2}\d u \, .
\ee
Now we introduce other change of variables
$t=\f{u}{\sigma}$, and  then the integral takes the form
\be
\t V_\pm=\f{\kappa \sigma}{4\sqrt{2}\alpha}
e^{\pm p\alpha} e^{\alpha^2/2\sigma^2}\int_{-z}^\infty
\left[2\sigma^2 t^2+4\sqrt{2}k_0\sigma t+4k_0^2 +\alpha^2 \right]
e^{-t^2/}\d t\, ,
\ee
with $z=x \pm iy=\f{\sqrt 2 k_0}{\sigma}\pm i\f{\alpha}{\sigma \sqrt 2}$. The integral is
\bas
\t I_\pm
&=&\int_{-\infty}^\infty
\left(2\sigma^2 t^2+4\sqrt{2}k_0\sigma t+4k_0^2 +\alpha^2 \right)
e^{-t^2}\d t -
\int_{-\infty}^{-z}
\left(2\sigma^2 t^2+4\sqrt{2}k_0\sigma t+4k_0^2 +\alpha^2 \right)
e^{-t^2}\d t \\
&=& \sqrt \pi k_0\left[
4k_0 +\f{\sigma^2 + \alpha^2}{k_0}
-\f{z\sigma^2 }{\sqrt\pi}e^{-z^2}-\f{\sigma^2}{2\sqrt\pi k_0}erfc(z)
-2\sqrt{2}\sigma \f{e^{-z^2}}{\sqrt\pi}
-\left( 2k_0 +\f{\alpha^2}{2k_0}\right) erfc(z) \right]\, .
\eas
Then $\t V_\pm$ takes the form
\bas
\t V_\pm &=&\f{\kappa }{8\alpha}
e^{\pm p\alpha} e^{\alpha^2/2\sigma^2} \sqrt{2\pi}\sigma k_0
\left[
4k_0 +\f{\sigma^2 + \alpha^2}{k_0} \right. \\
& &\hspace{3.5cm} \left.
-\f{z\sigma^2 }{\sqrt\pi}e^{-z^2}-\f{\sigma^2}{2\sqrt\pi k_0}erfc(z)
-2\sqrt{2}\sigma \f{e^{-z^2}}{\sqrt\pi}
-\left( 2k_0 +\f{\alpha^2}{2k_0}\right) erfc(z) \right]\, .
\eas
If we normalize using Eq. \eqref{norm0-ap} then
\bas
 V_\pm &=&\f{\kappa }{8\alpha}
e^{\pm p\alpha} e^{\alpha^2/2\sigma^2}
\left[
4k_0 +\f{\sigma^2 + \alpha^2}{k_0} \right. \\
& &\hspace{3cm} \left.
-\f{z\sigma^2 }{\sqrt\pi}e^{-z^2}-\f{\sigma^2}{2\sqrt\pi k_0}erfc(z)
-2\sqrt{2}\sigma \f{e^{-z^2}}{\sqrt\pi}
-\left( 2k_0 +\f{\alpha^2}{2k_0}\right) erfc(z) \right]
\eas
The values that we used in the manuscript was
\be \label{Vpm_02}
V_\pm
\approx \f{\kappa k_0}{2\alpha} e^{\pm p\alpha} e^{\alpha^2/2\sigma^2}
\left[1 +\f{\sigma^2 +\alpha^2}{4k_0^2} \right] \, .
\ee
%
Then the
error in the integration is
\bas
{\rm Error}( V_\pm) &=&-\f{\kappa }{8\alpha}
e^{\pm p\alpha} e^{\alpha^2/2\sigma^2}
\left[\f{e^{-z^2}}{\sqrt\pi} (z\sigma^2  +2\sqrt{2}\sigma)
 +erfc(z)\left( \f{\sigma^2}{2\sqrt\pi k_0}+2k_0 +\f{\alpha^2}{2k_0}\right)  \right]
\eas
Using the triangular inequality
\bas
|{\rm Error}( V_\pm)| &\leq & \f{\kappa }{8\alpha}
e^{\pm p\alpha} e^{\alpha^2/2\sigma^2}
\left[\f{e^{-|z|^2}}{\sqrt\pi} (|z|\sigma^2  +2\sqrt{2}\sigma)
 +|erfc(z)|\left( \f{\sigma^2}{2\sqrt\pi k_0}+2k_0 +\f{\alpha^2}{2k_0}\right)  \right]
\eas
Using the relation \eqref{erfc_approx_complex} 
(where $|$arg$z|=\arctan\f{y}{x}=\arctan\f{\alpha}{k_0} <\f{\pi}{4}$ because $\alpha\ll k_0$)
\bas
|{\rm Error}( V_\pm)| 
& < &\f{\kappa }{8\sqrt\pi \alpha}
e^{\pm p\alpha} e^{\alpha^2/2\sigma^2}e^{-|z|^2}
\left[ |z|\sigma^2  +2\sqrt{2}\sigma
 + \f{\sigma^2}{2\sqrt\pi k_0 |z|}+\f{2k_0}{|z|} +\f{\alpha^2}{2k_0|z|}  \right]\\
& < &\f{\kappa}{8\sqrt{\pi} \alpha} 2\sqrt 2 \sigma^2 \alpha^2 k_0 |z|
e^{\pm p\alpha} e^{\alpha^2/2\sigma^2}e^{-|z|^2}
\left[ \f{1}{2\sqrt 2 \alpha^2 k_0}
+\f{1}{\alpha^2  \sigma k_0 |z|} \right. \\
& & \left. + \f{1}{4\sqrt{2\pi} k_0^2 |z|^2 \alpha^2 }
 +\f{1}{|z|^2\sqrt 2 \sigma^2 \alpha^2 } 
 +\f{1}{4\sqrt 2 k_0^2|z|^2 \sigma^2}  \right]
\eas
As all quantities in the brackets are positive and less than one, we can bound the error as
\bas
|{\rm Error}( V_\pm)| 
& < &\f{5 \kappa \sigma^2 \alpha k_0 |z|}{2\sqrt{2\pi}} 
e^{\pm p\alpha} e^{\alpha^2/2\sigma^2}e^{-|z|^2}
\eas
Given that $V_\pm$ can be too small or large then the quantity that really gives a measure of the
error  is  the relative error $\frac{|{\rm Error}( V_\pm)| }{V_\pm}$. Using Eq. \eqref{Vpm_02}, this becomes
\bas
\frac{|{\rm Error}( V_\pm)| }{V_\pm}
& < &\f{5\alpha^2\sigma^2 |z|}{\sqrt{2\pi}}
e^{-|z|^2} \left(1+\f{\sigma^2+\alpha^2}{4k_0^2} \right)^{-1} \\
& < &\f{5\alpha^2 k_0\sigma}{\sqrt{2\pi}}
e^{-2k_0^2/\sigma^2} e^{-\alpha^2/2\sigma^2}
\f{\left(1+\f{\alpha^2}{4k_0^2} \right)^{1/2}}{\left(1+\f{\sigma^2+\alpha^2}{4k_0^2} \right)}
\eas
The relation between the parentheses is less than one, then
\bas
\frac{|{\rm Error}( V_\pm)| }{V_\pm}
& < &\f{5\alpha^2 k_0\sigma}{\sqrt{2\pi}}
e^{-2k_0^2/\sigma^2} e^{-\alpha^2/2\sigma^2}
\eas
Since $\sigma \gg \alpha$, then $e^{-\alpha^2/2\sigma^2}\approx 1$. Furthermore, as $k_0 \gg\sigma\gg\alpha$, then 
the negative exponential wins over the polynomial factor, which show that the error is small.
These errors were also studied numerically for $n>0$ and present similar behavior.

\section{Tables}
\label{app:c}

In this appendix we summarize the main quantities that were found
in the main text, for the generalized Gaussian states for $n=1$
and compare them to the $n=0$ case. It should be noted, that for higher
values of $n$, the structure of the terms is similar, with terms of the form
$\left(\f{\alpha}{k_0}\right)^m$ and $\left(\f{\sigma}{k_0}\right)^l$.

\begin{table}[h]
\begin{center}
\renewcommand{\arraystretch}{1.8}
\begin{tabular}{|c|c|c|}
\hline
& {\bf n=0} & {\bf n=1} \\
\hline $\|\t\chi\|_{\rm phy}^2$ & $ \sqrt{2\pi} k_0\sigma $
& $ \sqrt{2\pi} k_0\sigma\left( k_0^2+\f{3}{4}\sigma^2\right)  $  \\
\hline $V_\pm$ & $\alpha e^{\pm p\alpha} e^{\alpha^2/2\sigma^2}
\left(\f{1}{2}k_0 +\f{\sigma^2 +\alpha^2}{8k_0} \right)$ & $\alpha
e^{\pm p\alpha} e^{\alpha^2/2\sigma^2}
\left(\f{3\sigma^4+24k_0^2\sigma^2+2\alpha^2\sigma^2
+16k_0^4 +8k_0^2\alpha^2 +\alpha^4}{24k_0\sigma^2+32k_0^3}\right) $ \\[2mm]
\hline $W_0$ & $\alpha_2 \left(k_0^2 +\f{3\sigma^2}{4}\right) $ &
$ \alpha_2
\left(\f{15\sigma^4+40k_0^2\sigma^2+16k_0^4}{12\sigma^2+16k_0^2}\right)$ \\[2mm]
\hline $W_\pm $ & $\alpha_3 e^{\pm 2p\alpha}e^{2\alpha^{2}/\sigma^2}
 \left(k_0^2 +\f{3}{4}\sigma^2+\alpha^2 \right)  $
& $\alpha_3 e^{\pm 2p\alpha}e^{2\alpha^{2}/\sigma^2}
\left(\f{15\sigma^4+40k_0^2\sigma^2+24\alpha^2\sigma^2+16k_0^4+32\alpha^2k_0^2+16\alpha^4}
{12\sigma^2+16k_0^2}\right) $ \\[2mm]
\hline $\langle \hat p_\phi \rangle $ & $\hbar k_0\left(1
+\f{\sigma^2}{4 k_0^2} \right)$ & $\hbar
k_0\f{\left(1+\f{3\sigma^2}{2k_0^2}+\f{3\sigma^4}{16k_0^4}\right)}
{\left(1+\f{3\sigma^2}{4k_0^2}\right)} $ \\[4mm]
\hline $\langle \hat p_\phi^2 \rangle $ & $\hbar^2 k_0^2\left(1
+\f{3\sigma^2}{4 k_0^2} \right) $ & $\hbar^2k_0^2\f{\left(1+
\f{5\sigma^2}{2k_0^2}+\f{15\sigma^4}{16k_0^4}\right)}
{\left(1+\f{3\sigma^2}{4k_0^2}\right)} $ \\[4mm]
\hline $(\Delta p_\phi)^2 $ & $\f{\hbar^2\sigma^2}{4}\left( 1
-\f{\sigma^2}{4 k_0^2} \right) $ & $\f{\hbar^2\sigma^2}{4}
\f{\left(1+
\f{3\sigma^2}{4k_0^2}+\f{9\sigma^4}{16k_0^4}-\f{9\sigma^6}{64k_0^6}\right)}
{\left(1+\f{3\sigma^2}{2k_0^2}+\f{9\sigma^4}{16k_0^4}\right)} $ \\[4mm]
\hline $V_{\rm bounce} $ & $\alpha e^{\alpha^2/2\sigma^2} k_0
\left(1 +\f{\sigma^2 +\alpha^2}{4k_0^2} \right) $ & $\alpha
e^{\alpha^2/2\sigma^2}k_0\left(
\f{1+\f{3\sigma^4}{16k_0^4}+\f{3\sigma^2}{2k_0^2}+\f{\alpha^2\sigma^2}{8k_0^4}
 +\f{\alpha^2}{2k_0^2} +\f{\alpha^4}{16k_0^4}}
{1+\f{3\sigma^2}{4k_0^2}}\right) $ \\[3mm] \hline
$\f{(\Delta V)^2}{\langle V \rangle^2} $ & $e^{\alpha^{2}/\sigma^2}
\f{\left(1 +\f{3\sigma^2}{4k_0^2}+\f{\alpha^2}{k_0^2} \right)}
{\left(1 +\f{\sigma^2 +\alpha^2}{4k_0^2} \right)^2} -1 $ &
$e^{\alpha^{2}/\sigma^2}\left(1+\f{3\sigma^2}{4k_0^2}\right)
\f{\left({1+\f{15\sigma^4}{16k_0^4}+\f{5\sigma^2}{2k_0^2}
+\f{3\alpha^2\sigma^2}{2k_0^4}
+\f{2\alpha^2}{k_0^2}+\f{\alpha^4}{k_0^4}}\right)}
{\left({1+\f{3\sigma^4}{16k_0^4}+\f{3\sigma^2}{2k_0^2}
+\f{\alpha^2\sigma^2}{8k_0^4}
+\f{\alpha^2}{2k_0^2} +\f{\alpha^4}{16k_0^4}}\right)^2}-1 $ \\[5mm] \hline
\end{tabular}
\end{center}
\vspace{-0.6cm}
\end{table}

\end{appendix}

\end{document}